\documentclass[aps,prl,reprint,superscriptaddress,longbibliography]{revtex4-2}
\usepackage{amsmath}
\usepackage{graphicx}   
\graphicspath{{./}} 
\usepackage{amssymb}
\usepackage{xcolor}
\usepackage{bm}
\usepackage{hyperref}

\begin{document}

\title{Emergent Trion Resonance Driven by Lattice Reconstruction in a Moir\'e Superlattice}

\author{Zhida Liu}
\thanks{These authors contributed equally to this work.}
\affiliation{Department of Physics and Center for Complex Quantum Systems, The University of Texas at Austin, Austin, Texas, 78712, USA.}
\author{Haonan Wang}
\thanks{These authors contributed equally to this work.}
\affiliation{Department of Physics, Washington University in St Louis, St. Louis, MO, 63136, USA.}
\author{Xiaohui Liu}
\affiliation{Department of Physics and Center for Complex Quantum Systems, The University of Texas at Austin, Austin, Texas, 78712, USA.}
\affiliation{Center for Dynamics and Control of Materials and Texas Materials Institute, The University of Texas at Austin, Austin, Texas, 78712, USA.}
\author{Yue Ni}
\affiliation{Department of Physics and Center for Complex Quantum Systems, The University of Texas at Austin, Austin, Texas, 78712, USA.}
\author{Hongtao Yan}
\affiliation{Department of Physics and Center for Complex Quantum Systems, The University of Texas at Austin, Austin, Texas, 78712, USA.}
\affiliation{Center for Dynamics and Control of Materials and Texas Materials Institute, The University of Texas at Austin, Austin, Texas, 78712, USA.}
\author{Frank Y. Gao}
\affiliation{Department of Physics and Center for Complex Quantum Systems, The University of Texas at Austin, Austin, Texas, 78712, USA.}
\affiliation{Center for Dynamics and Control of Materials and Texas Materials Institute, The University of Texas at Austin, Austin, Texas, 78712, USA.}
\author{Saba Arash}
\affiliation{Department of Physics and Center for Complex Quantum Systems, The University of Texas at Austin, Austin, Texas, 78712, USA.}
\affiliation{Center for Dynamics and Control of Materials and Texas Materials Institute, The University of Texas at Austin, Austin, Texas, 78712, USA.}
\author{Hyunsue Kim}
\affiliation{Department of Physics and Center for Complex Quantum Systems, The University of Texas at Austin, Austin, Texas, 78712, USA.}
\author{Dong Seob Kim}
\affiliation{Department of Physics and Center for Complex Quantum Systems, The University of Texas at Austin, Austin, Texas, 78712, USA.}
\author{Xiangcheng Liu}
\affiliation{Department of Physics and Center for Complex Quantum Systems, The University of Texas at Austin, Austin, Texas, 78712, USA.}
\author{Xiaoxiao Yu}
\affiliation{Department of Physics and Center for Complex Quantum Systems, The University of Texas at Austin, Austin, Texas, 78712, USA.}
\author{Yongxin Zeng}
\affiliation{Department of Physics and Center for Complex Quantum Systems, The University of Texas at Austin, Austin, Texas, 78712, USA.}
\author{Jiamin Quan}
\affiliation{Department of Physics and Center for Complex Quantum Systems, The University of Texas at Austin, Austin, Texas, 78712, USA.}
\author{Di Huang}
\affiliation{Department of Physics and Center for Complex Quantum Systems, The University of Texas at Austin, Austin, Texas, 78712, USA.}
\author{Kenji Watanabe}
\affiliation{Research Center for Electronic and Optical Materials, National Institute for Materials Science, 1-1 Namiki, Tsukuba 305-0044, Japan.}
\author{Takashi Taniguchi}
\affiliation{Research Center for Materials Nanoarchitectonics, National Institute for Materials Science,  1-1 Namiki, Tsukuba 305-0044, Japan.}
\author{Edoardo Baldini}
\affiliation{Department of Physics and Center for Complex Quantum Systems, The University of Texas at Austin, Austin, Texas, 78712, USA.}
\affiliation{Center for Dynamics and Control of Materials and Texas Materials Institute, The University of Texas at Austin, Austin, Texas, 78712, USA.}
\author{Keji Lai}
\affiliation{Department of Physics and Center for Complex Quantum Systems, The University of Texas at Austin, Austin, Texas, 78712, USA.}
\affiliation{Center for Dynamics and Control of Materials and Texas Materials Institute, The University of Texas at Austin, Austin, Texas, 78712, USA.}
\author{Allan H. MacDonald}
\affiliation{Department of Physics and Center for Complex Quantum Systems, The University of Texas at Austin, Austin, Texas, 78712, USA.}
\affiliation{Center for Dynamics and Control of Materials and Texas Materials Institute, The University of Texas at Austin, Austin, Texas, 78712, USA.}
\author{Chih-Kang Shih}
\affiliation{Department of Physics and Center for Complex Quantum Systems, The University of Texas at Austin, Austin, Texas, 78712, USA.}
\affiliation{Center for Dynamics and Control of Materials and Texas Materials Institute, The University of Texas at Austin, Austin, Texas, 78712, USA.}
\author{Jamie Warner}
\affiliation{Walker Department of Mechanical Engineering and the Texas Materials Institute, The University of
Texas at Austin, Austin, Texas 78712, USA. }
\author{Li Yang}
\affiliation{Department of Physics, Washington University in St Louis, St. Louis, MO, 63136, USA.}
\author{Xiaoqin Li}
\email{elaineli@physics.utexas.edu}
\affiliation{Department of Physics and Center for Complex Quantum Systems, The University of Texas at Austin, Austin, Texas, 78712, USA.}
\affiliation{Center for Dynamics and Control of Materials and Texas Materials Institute, The University of Texas at Austin, Austin, Texas, 78712, USA.}
\thanks{Corresponding author}

\begin{abstract}

 We investigate how many-electron excited states emerge in twisted MoSe$_2$ homobilayers when the lattice reconstructions evolve. Notably, we identify a new trion resonance that arises in the transition regime of lattice reconstruction, where gradual changes in atomic alignment between the layers occur. Magnetic field–dependent measurements, supported by first-principles calculations, indicate that the exciton forms at the K valley while the doped hole resides in the $\Gamma$-valley. First-principles calculations further indicate that two nearly degenerate exciton resonances can arise, localized at different sites within the moiré supercell. 
 We propose that the new trion resonance is a ``charge-transfer" trion, in which the electron–hole pair is spatially separated from the doped hole. The emergence of these complex excited states stems from the distinct moiré potentials acting on holes and excitons, resulting in their different spatial distribution within the superlattice. 

\end{abstract}
\maketitle
Semiconductor moir\'e superlattices formed by stacking two atomically thin layers have been established as a versatile platform to realize correlated electronic phases~\cite{wang2020correlated,tang2020simulation,shimazaki2020strongly,regan2020mott} and novel optical properties~\cite{regan2022emerging,huang2022excitons,andersen2021excitons}. A variety of exciton resonances including intra-, interlayer, and hybrid excitons dominate the optical spectra of transition metal dichalcogenide (TMD) bilayers. 
The resonant energy, dynamics, and nonlinear response of various excitonic resonances~\cite{tran2019evidence, seyler2019signatures,jin2019observation,alexeev2019resonantly,zhang2018moire} are controllable by the twist angle between the layers~\cite{merkl2020twist,choi2021twist,zhang2020twist,lin2024moire}. Upon doping, the ground state may evolve into correlated insulators or Wigner crystal phases ~\cite{regan2020mott, xiong2023correlated, xu2020correlated}, while new excited states such as trions and attractive or repulsive polarons emerge. The microscopic nature of many-body excited states in a given superlattice remains poorly understood.

 The vast majority of previous studies have treated excitons as composite particles moving in a smooth moir\'e potential~\cite{wu2017topological,brem2020tunable}. This assumption should be scrutinized. Because the conduction- and valence-band extrema may occur at different sites within a supercell~\cite{zhang2020flat,li2021imagingdischarge,weston2020atomic}, many-body states with intricate internal structure can emerge~\cite{zeng2022strong}. A prominent example is the recently discovered “charge-transfer” exciton, in which the electron and hole localize at different supercell sites. These excitons have been identified in WSe$_2$/WS$_2$ bilayers through comparisons between the first-principles calculations and doping- and field-dependent optical spectra~\cite{naik2022intralayer}, and in WS$_2$ bilayers using atomic-resolution photocurrent imaging~\cite{li2024imaging}. 
 
Here, we report the observation of a new type of trion resonance in twisted MoSe$_2$ bilayers driven by lattice reconstruction.  We compare the optical spectra of several MoSe$_2$ bilayers near $H$-stacking, with twist angles of 60$^{\circ}$, 59.6$^{\circ}$, 59.2$^{\circ}$, and 57.5$^{\circ}$.  
In the 57.5$^{\circ}$ bilayer, scanning transmission electron microscopy (STEM) reveals gradual variations in atomic alignment, placing the structure in the transition regime, an assignment further corroborated by high-resolution Raman spectroscopy. Whereas other hole-doped bilayers exhibit only a single trion resonance, the 57.5$^{\circ}$  bilayer shows an additional trion resonance. By analyzing the Zeeman splitting of the resonances extracted from magnetic field-dependent measurements and performing first-principles calculations, we assign the doped hole to the $\Gamma$ valley in the valence band. Calculations further suggest that two nearly degenerate exciton resonances can form with different wavefunction distributions within the supercell. We propose that the newly identified trion resonance in the 57.5$^{\circ}$ bilayer is a novel ``charge-transfer" trion, in which the optically created electron-hole pair is spatially separated from the doped hole. These complex excited states arise from the distinct moir\'e potentials experienced by holes and excitons, which in turn lead to different spatial distributions within the supercell.
 
Natural and twisted MoSe$_2$ bilayers encapsulated between hBN layers are incorporated in dual-gate devices (Fig.~\ref{fig:fig1}a with additional details in the supplementary material~\cite{SM}), allowing us to tune the doping level and an out-of-plane electric field independently. All data presented in this paper are taken at zero electric field and a temperature of $\sim$ 4.7 K unless stated otherwise. In natural MoSe$_2$ bilayers ($\theta$ = 60$^{\circ}$), the energetically favorable stacking is the $AA'$ stacking, where the centers of the hexagons in two layers coincide and the Mo atoms in the top layer are vertically aligned with the Se atoms in the bottom layer. For a moderate twist angle~(i.e., 57.5$^\circ$), there are gradual atomic alignment variations between the two layers, as simulated in Fig.~\ref{fig:fig1}b. In addition to the $AA'$ stacking, two other high-symmetry stackings $A'B$ and $AB'$ retain the three-fold rotational symmetry. At the $A'B$ ($AB'$) points,~the Se (Mo) atoms in the two layers are vertically aligned. Colored dots mark the high symmetry points.

In moir\'e superlattices with a small twist angle from the commensurate $H$- and $R$-stacking, lattice reconstructions driven by the competition between interlayer coupling and intrinsic strain can be described in three regimes~\cite{carr2020electronic, quan2021phonon}: fully relaxed regime, transition regime, and rigid lattice regime. We perform first-principles calculations to predict the effect of lattice reconstructions in twisted MoSe$_2$ bilayers.  While the area of AB' (A'B) stacking significantly shrinks in a $59.2^{\circ}$ bilayer as shown in Fig.~S2, it is largely preserved in the $57.5^{\circ}$ bilayer. Colored dots in Fig.~\ref{fig:fig1}b indicate the high-symmetry points. These calculations are consistent with previous studies of moir\'e superlattices~\cite{weston2020atomic,naik2022intralayer} (section 3 in SI). 

The atomic-resolution STEM image (Fig.~\ref{fig:fig1}c) confirms the expected gradual variations in atomic alignment within the 57.5$^\circ$ MoSe$_2$ bilayer in the transition regime, corresponding to brightness modulations. The twist-angle precision and the uniformity of the moiré pattern are further validated by a high-angle annular dark-field (HAADF) STEM image (Fig.~S3).  To ensure that the 57.5$^\circ$ bilayer used in optical measurements retains these alignment variations, we performed high-resolution Raman spectroscopy measurements with a 532 nm excitation laser from the MoSe$_2$ bilayers. The results are shown in Fig.~\ref{fig:fig1}d. In the low-frequency range, only one shear (S) mode is observed in natural bilayers ($\theta$ = 60$^\circ$) and fully relaxed lattices ( $\theta$~=~59.2$^\circ$ and 59.6$^\circ$)~\cite{weston2020atomic}. In contrast, the emergence of the layer breathing (LB) mode in the $\theta = 57.5^\circ$ bilayer suggests that gradual atomic variations occur in this sample~\cite{puretzky2016twisted}. The varying atomic alignments between the two layers, confirmed by the combination of STEM and Raman spectra, are a critical condition for the formation of the new charge transfer trion.\\ 
 \begin{figure}
\centering
    \includegraphics[width=8cm]{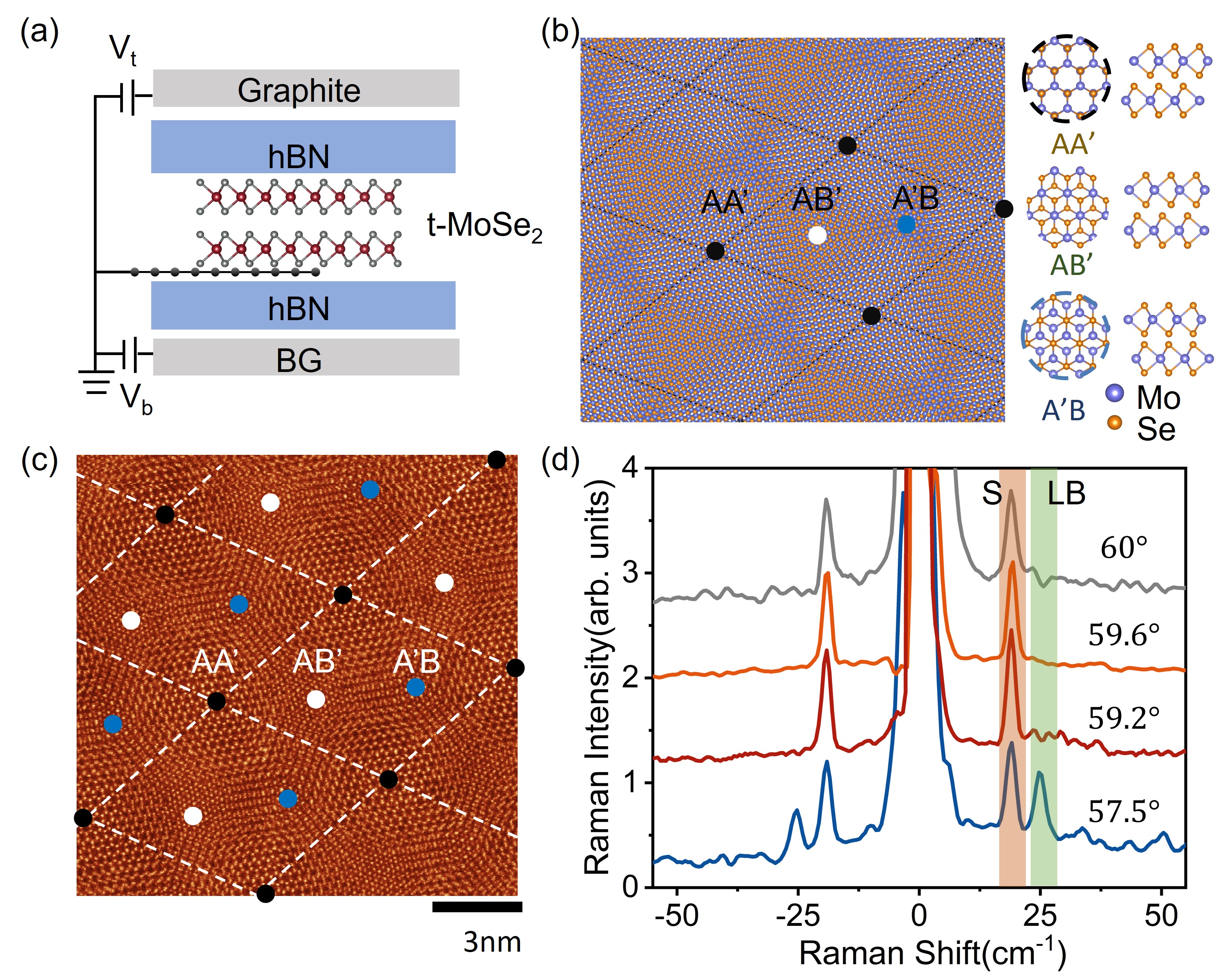}
   \caption{\textbf{
    Schematics of MoSe$_2$ bilayer devices and atomic alignment variations confirmed by TEM and Raman spectra.}
    \textbf{(a)} Dual-gate device used to control the doping density in MoSe$_2$ bilayers. 
    \textbf{(b)} Calculated atomic structure of 57.5$^\circ$ twisted MoSe$_2$ bilayers. The black dashed lines outline the supercells. Three high symmetry points are labeled as $AA'$,~$AB'$, and~$A'B$.
    \textbf{(c)} TEM images of a 57.5$^\circ$ MoSe$_2$ bilayer. The high symmetry points are marked by colored dots.
    \textbf{(d)} Raman spectra from four bilayers with $\theta$ of 60$^\circ$ (grey),~{59.6$^\circ$ (orange)},~{59.2$^\circ$ (red)}, and 57.5$^\circ$ (blue), respectively. A shear mode (S), indicated by the orange stripe, is present in all three bilayers, while the layer breathing (LB) mode, indicated by the green stripe, is only observable in the 57.5$^\circ$ bilayer.
    }  
    \label{fig:fig1}
    \end{figure}
    
\begin{figure}[t]
\includegraphics[width=9cm]{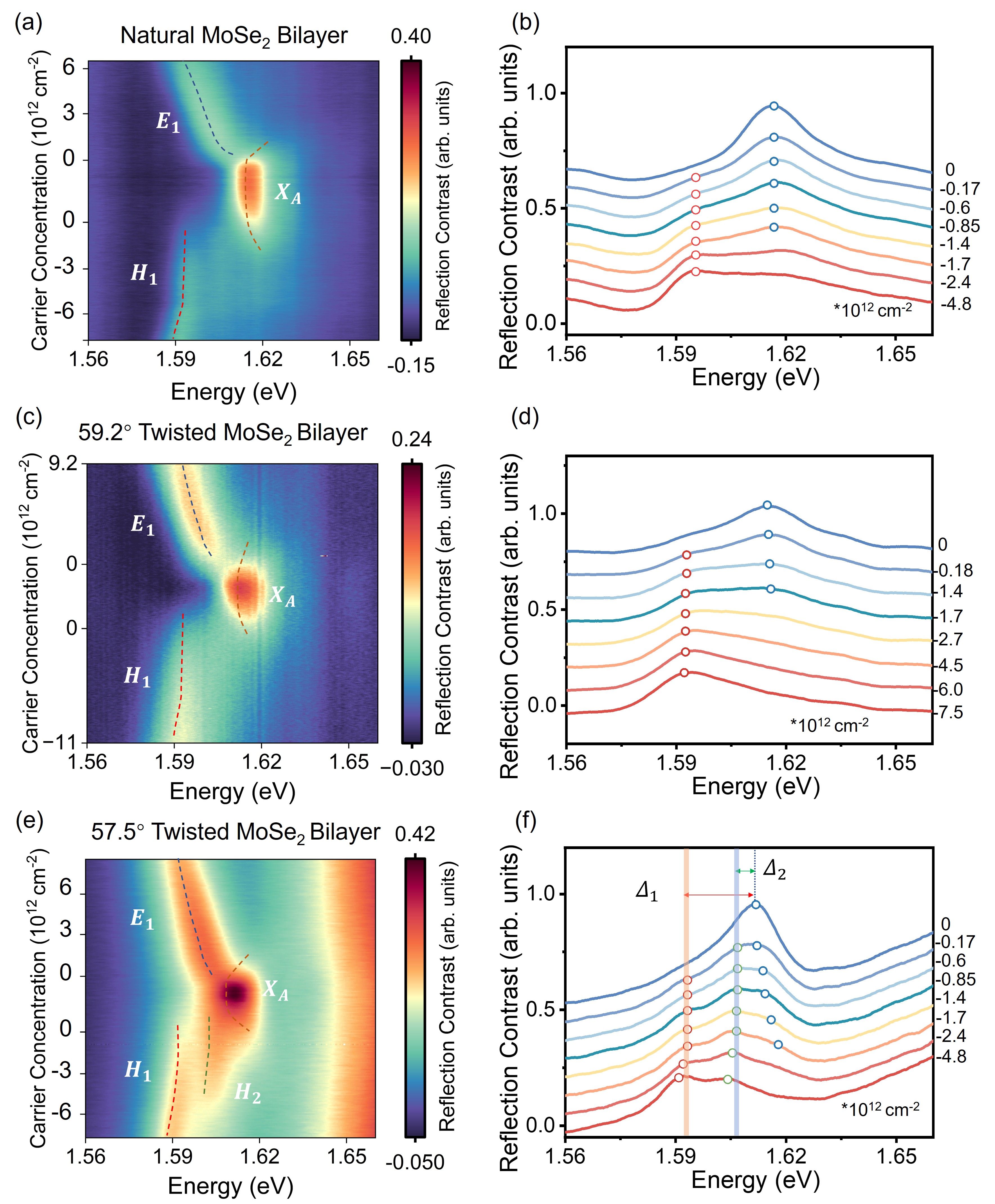}
\centering
    \caption{\textbf{Doping dependent differential reflectivity in three different bilayers. } 
    Differential reflectivity spectra of  \textbf{(a)} a natural bilayer, {\textbf{(c)} a fully relaxed 59.2$^\circ$, }and \textbf{(e)} a 57.5$^\circ$ bilayer. In both \textbf{(a)} and \textbf{(c)}, the A exciton resonance in the intrinsic regime evolves into two branches with electron doping. The lower energy $E_1$ resonance red-shifts with increasing electron doping. With hole doping, a bound state $H_1$ forms and only red-shifts beyond a certain hole doping density. 
    In \textbf{(e)} a new trion resonance $H_2$ appears. Dashed lines serve as a guide to the eye, highlighting the energy shifts of different resonances. Linecuts of reflectivity spectra taken at several doping densities in \textbf{(b)} the natural bilayer, \textbf{(d)} the 59.2$^\circ$, and \textbf{(f)} the 57.5 $^\circ$  bilayer. $\Delta _{1}$ ($\Delta _{2}$) corresponds to the binding energy of $H_1$ ($H_2$).}
    \label{fig:fig2}    
\end{figure}

We compare the optical reflectivity spectra from the natural bilayer, 59.2$^\circ$, and 57.5$^\circ$ twisted bilayers. The two-dimensional (2D) contours as a function of doping density and energy are shown in Fig.~\ref{fig:fig2}~(a,c,e), respectively. The relatively broad linewidths of the exciton and trion resonances are partially due to additional decay pathways due to phonon-assisted inter-valley scattering ~\cite{helmrich2021phonon}. When electron doping increases in all three bilayers, the $A$ exciton~($X_A$) evolves into two branches, exhibiting either a continuous red-shift ($E_1$) or a blue-shift. These resonances are attributed to the formation of attractive and repulsive Fermi polarons~\cite{efimkin2021electron,sidler2017fermi}. While doped electrons are not fully delocalized, their spatial extent is not negligible compared to the supercell size, which is consistent with the calculated dispersive conduction band edge states. On the hole doping side, in addition to a blue-shift resonance from the A exciton, a positive trion ($H_1$) is observed. Doped holes are strongly localized in real space due to the flat valence bands predicted by first-principles calculations~\cite{naik2020origin}. Polaron and trion theories converge at low doping density. For simplicity, we will use trions to refer to emerging optical resonances in hole-doped bilayers in this manuscript. In the case of $H_1$ in the natural and 59.2$^\circ$ twisted bilayer, the binding energy is approximately 25~meV~\cite{gao2016dynamical}. Similar results are shown in 59.6 $^\circ$ device in Fig.~S4. In the $57.5^\circ$ twisted bilayer, a new trion resonance $H_2$ with a smaller binding energy of $\sim$ 12 meV appears (similar data in another device shown in Fig.~S5b). Multiple horizontal linecuts taken at several doping densities in all three bilayers are displayed in Fig.~\ref{fig:fig2}(b,d,f), respectively. The peak energies of the exciton and trion resonances are highlighted by circles and extracted through multi-Lorentzian fitting (Section 4 in SI and Fig.~S5c). Spectra taken from two 57.5$^\circ$ samples (see SI for details) are reproducible. Exciton and trion resonances broaden with temperature, making them difficult to resolve above 45 K (Data included in the SI). The reflectivity and Raman spectra are mutually consistent with our hypothesis that H$_2$ only exists in superlattices beyond the fully relaxed regime, although we cannot completely rule out disorder or a reduced oscillator strength in the 59.2$^\circ$ and 59.6$^\circ$ samples as potential reasons for the absence of the H$_2$ resonance. Additional experiments are needed to precisely define the twist-angle window over which H$_2$ can exist.

\begin{figure}
 \includegraphics[width=9cm]{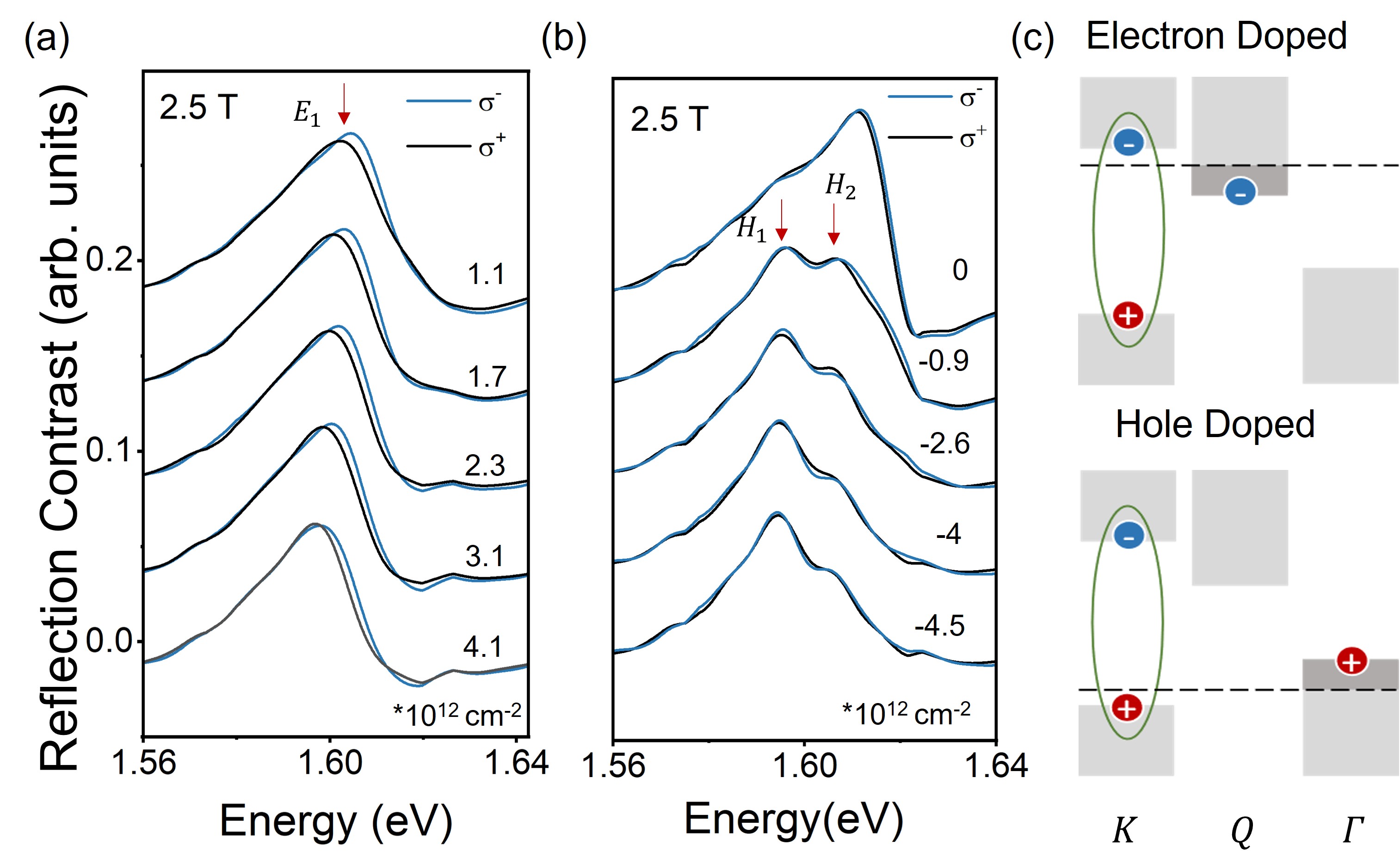}
    \caption{\textbf{Band structure calculations and trion configurations in momentum space.}
    {\textbf{(a,b)} Helicity-resolved reflectivity linecuts taken in the presence of 2.5~T magnetic field at several different  \textbf{(a)} electron and \textbf{(b)} hole doping densities.}
    \textbf{(c)} Trion configurations in momentum space. The top panel shows that doped electrons reside at the Q valleys of the first CBM. The bottom panel shows that doped holes first occupy the $\Gamma$ valleys at the first VBM. 
      }
    
    \label{fig:fig3}    
\end{figure}

Helicity-resolved magneto-optical reflectivity measurements further reveal the differences between many-body excited states between the electron- and hole-doped bilayers. Several spectra taken at 2.5 T and at various electron and hole doping densities are displayed in Fig.~\ref{fig:fig3}a and b, respectively. The electron-doped spectra exhibit a significant Zeeman splitting associated with spin-polarized electrons at the $Q$ valleys of the conduction band~\cite{forste2020exciton}. Fitting the spectra with Lorentz functions and using a simple relation~$\Delta$E =~g$\mu_B$B, where $\Delta$E is the energy difference between $\sigma^-$ and $\sigma^+$ excitations, we extract the absolute value of a g-factor of 20~$\pm1.4$ at 1.1$\times 10^{12}~cm^{-2}$~\cite{smolenski2019interaction}. In contrast, spectra taken at several hole doping densities at 2.5~T reveal minimal spectral shift for both $H_1$ and $H_2$. Similarly small Zeeman splitting of trions has been observed in twisted WSe$_2$/MoSe$_2$ trilayers~\cite{campbell2024interplay}. We speculate that the Zeeman splitting is much smaller than the trion linewidth for the positive trions involving holes at the $\Gamma$ valley with spin degeneracy as illustrated in Fig.~\ref{fig:fig3}c. Additional evidence supporting doped holes in the $\Gamma$ valley is the redshift of $H_1$ and $H_2$ with increasing doping density~(Fig.~2c and Fig.~S6). This red shift is opposite to the blue shift of the positively charged trions in monolayers (Fig.~S7), supporting our conclusion that the doped holes do not occupy the same valley in MoSe$_2$ monolayers and bilayers.

We also investigate how the spectra evolve with an applied electric field. While an E-field can shift the band alignment between $K$ and $\Gamma$ valleys of the valence bands in principle, no significant changes are observed in the reflectivity spectra at a constant hole doping of -0.8$\times 10^{12}~\text{cm}^{-2}$ (Fig.~S8). The large energy difference between the $K$ and $\Gamma$ valleys likely requires a field strength beyond the accessible range. All spectroscopic features associated with doped holes in the $\Gamma$ valley are consistent with previous observations in MoSe$_2$/WSe$_2$ heterostructures~\cite{campbell2024interplay}. 
\begin{figure}[t]
\centering
 \includegraphics[width=8cm]{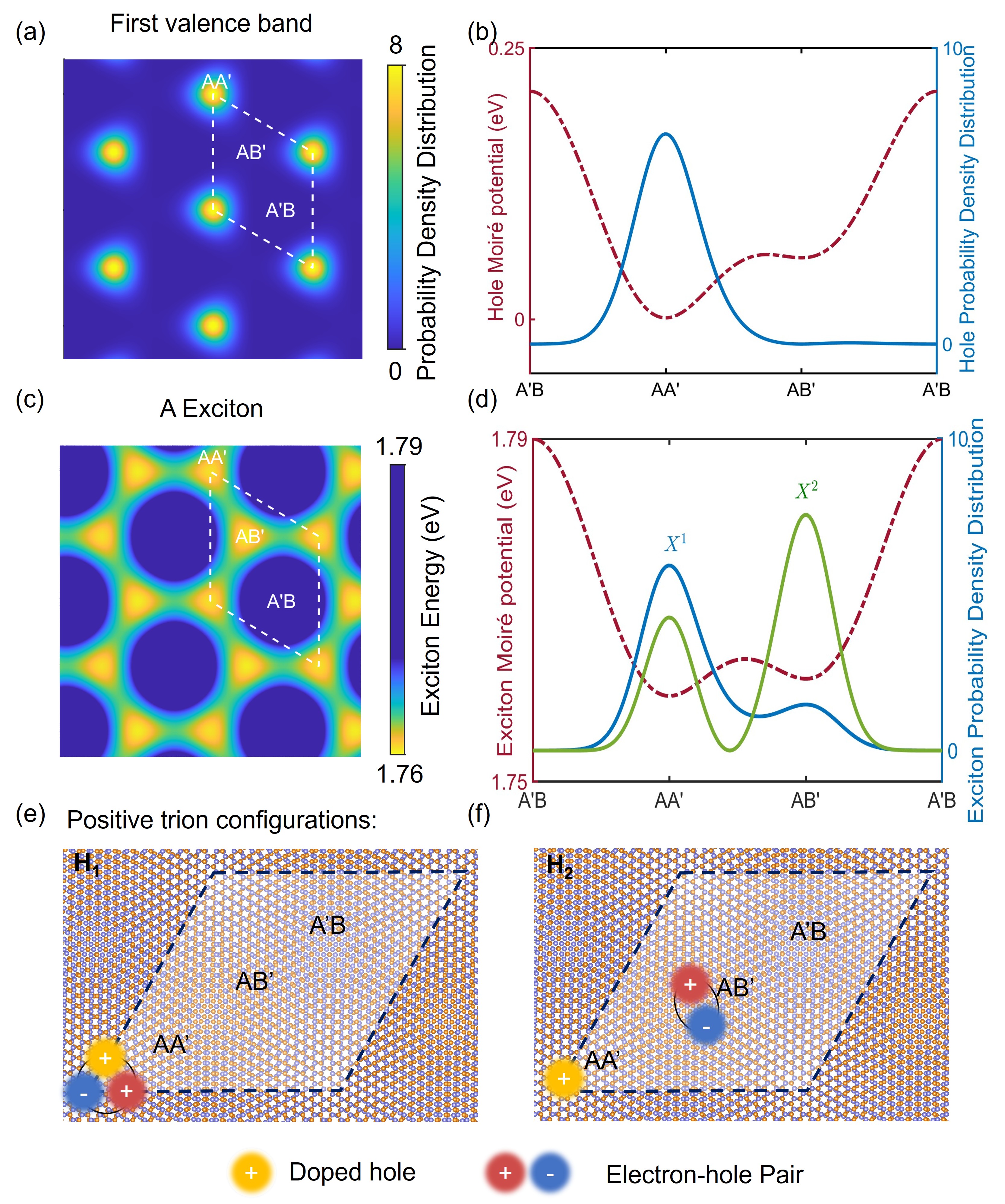}
    \caption{\textbf{Spatial distributions of hole states, excitons, trions.} \textbf{(a)} Probability density distribution of the first moir\'e valence band at the $\gamma$ point. \textbf{(b)} Hole moir\'e potential (red dash-dotted line) and probability density (blue solid line) of the first moir\'e valence miniband along the high-symmetry stacking line. The doped holes are mainly localized at the $AA'$ site. \textbf{(c)} Lowest exciton energy distributions where the exciton moir\'e potential only differs by $\sim$ 2 meV between the $AA'$ and $AB'$ sites. \textbf{(d)}~Exciton moir\'e potential (red dash-dotted line), probability density distribution of the lowest (blue solid line) and second lowest (green solid line) moir\'e exciton states along the high-symmetry stacking line. \textbf{(e,f)} Illustration of two types of positively charged trions where the extra hole is at the same (different) location as the electron-hole pair for $H_1$ ($H_2$).}
    \label{fig:fig4}    
\end{figure} 
The assumed distribution of dopants in momentum space is supported by first-principles GW simulations. The electronic self-energy significantly increases the density functional theory (DFT) band gap. For example, for the AA' stacking, the quasiparticle band gap is increased from 1.14 eV to 1.75 eV (Section 3 in SI), as expected~\cite{hybertsen1986electron}. More importantly, the quasiparticle conduction band minimum (CBM) shifts from the $K$ point to the $Q$ point, while the valence band maximum (VBM) remains at the $\Gamma$ point, as shown in Fig.~S9b. This calculation supports the hypothesis that doped holes (electrons) are primarily located at the $\Gamma$ ($Q$) point in momentum space, as illustrated in Fig.~3c. Consequently, a positive trion consists of a doped hole at the $\Gamma$ point coupled to an electron-hole pair generated via a $K$-$K$ optical transition, consistent with magnetic field-dependent spectra in Fig.~\ref{fig:fig3}a-b.  

To further elucidate the real-space distribution of doped electrons and holes, we directly calculate the electronic structure of the twisted structures. The quasiparticle moir\'e potential can be approximated through a first-order Fourier expansion
\begin{equation}   \label{eq1} 
\begin{aligned}     
    \Delta(\mathbf r)=V_0+V_1\sum_{j=1,3,5}\cos(\mathbf G_j\cdot \mathbf r+\psi),
	\end{aligned}
\end{equation}
where the summation runs over the three nearest-neighbor reciprocal lattices, and $\mathbf G$ denotes the reciprocal lattice vector of the moir\'e superlattice. The equation of motion of doped carriers is determined by the quasiparticle moir\'e Hamiltonian
\begin{equation}\label{eq2}    
\begin{aligned}     
\mathcal{H}&=\frac{\hbar^2 \mathbf Q^2}{2m^*}+\Delta(\mathbf r),
\end{aligned}
\end{equation}  
where the first term denotes the kinetic energy with $m^*$ representing the quasiparticle effective mass. All parameters in Eq. \eqref{eq2} can be determined from first-principles $GW$ calculations at three high-symmetry stacking ($AA^{\prime}$, $A^{\prime}B$, and $AB^{\prime}$). 

The interpolated moir\'e band structure for the $57.5^{\circ}$ bilayer is presented in Fig.~S9c, in which the highest valence band is flat, consistent with previous computational studies~\cite{naik2020origin,xian2021realization}. Fig.~\ref{fig:fig4}a shows the 2D contour plot of the charge density in the highest valence band, with the supercell indicated by the white dashed parallelogram. Fig.~\ref{fig:fig4}b (blue line) presents a line cut along the high-symmetry path ($A^{\prime}B$–$AA^{\prime}$–$AB^{\prime}$–$A^{\prime}B$). The hole probability density distribution exhibits a single peak, indicating that doped holes are predominantly localized in the $AA^{\prime}$ region. This localization results from the moiré potential (red dashed line in Fig.~\ref{fig:fig4}b) acting on the hole.

To understand why two types of trions can form, we calculate the two-particle excitation (exciton) spectrum in the 57.5$^{\circ}$ bilayer. The exciton Bohr radius is approximately 1–2 nm, which is significantly smaller than the moir\'e supercell size of $\sim$ 8 nm~\cite{qiu2013optical}.  Thus, we treat the lowest-energy exciton as a composite particle moving within a slowly varying exciton moir\'e potential. We first solve the Bethe-Salpeter Equation~\cite{rohlfing2000electron} at three high-symmetry stacking sites ($AA^{\prime}$, $A^{\prime}B$, and $AB^{\prime}$) to obtain exciton energies and dipole oscillator strengths (details in the SI). Taking into account the exciton energy, the exciton moir\'e potential is solved approximately using Eq.~\eqref{eq1}. The final step in finding the exciton spatial distribution is to solve Eq.~\eqref{eq2} with exciton moir\'e potential and effective mass. We plot how the exciton energy varies with the location of the center of mass wavefunction in the superlattice in Fig.~\ref{fig:fig4}c, in which bright yellow corresponds to low energy. AA' and AB' are two energetically favorable locations for exciton distribution. The same conclusion can be drawn from calculating the exciton moir\'e potential (red dashed line in Fig.~\ref{fig:fig4}d). Because the moir\'e potential experienced by holes and excitons differs, the doped holes and optically excited excitons may occupy either the same or different locations within a supercell, leading to tightly bound trions (H$_1$) or charge transfer trions (H$_2$).

Two excitons ($X^1$ and $X^2$) residing at the potential minima of $AA'$ and $AB'$ are nearly degenerate, with a small energy difference of 2 meV (Table S1 in the SI). Their center of mass wavefunction probability density distribution is plotted in (Fig.~\ref{fig:fig4}d). While $X^1$ primarily resides at $AA'$ sites, $X^2$ is more delocalized than $X^1$ and has a higher probability at $AB'$ sites.~These two exciton states are spectrally unresolved in experiments, i.e., only one resonance ($X_A$) is observed in the intrinsic~(undoped) regime (Fig.~\ref{fig:fig2}b). Their distinct spatial distribution, however, is clearly manifested when hole doping is introduced. Because doped holes always reside on the $AA'$ sites,~two types of trions ($H_1$ and $H_2$) form as illustrated in Fig.~\ref{fig:fig4}(e,f). In the case of the $H_1$ resonance, the exciton $X^1$ significantly overlaps with the doped hole, forming a tightly bound trion. In contrast, due to the relatively larger spatial separation between the doped hole and the exciton X$^2$, the charge transfer $H_2$ exhibits a smaller binding energy. In this qualitative discussion, we treat excitons and doped holes separately and the trion effect is not explicitly calculated due to the computational challenges.

We place our findings in the context of recent work on charge distributions within moir\'e supercells~\cite{naik2022intralayer,li2024imaging}. The first study of charge-transfer excitons in WSe$_2$/WS$_2$ bilayers, like the present work, combined optical spectroscopy with first-principles calculations~\cite{naik2022intralayer}. Direct imaging of the electron-hole separation was later achieved in twisted WS$_2$ bilayers using scanning tunneling microscopy~\cite{li2024imaging}, although this method cannot resolve the positions of electron-hole pairs within a supercell. In twisted WSe$_2$ homobilayers, exciton resonances were also found to vary with supercell size through combined scanning electron microscopy and optical spectroscopy in fully relaxed moir\'e superlattices with triangular domains~\cite{andersen2021excitons}. These variations were attributed to excitons occupying distinct regions of the supercell, a central premise that also underlies our work. For future studies, it would be interesting to image exciton wavefunctions in doped moir\'e superlattices by combining annular dark-field scanning transmission electron microscopy, low-loss electron energy-loss spectroscopy, and hyperspectral analysis under cryogenic conditions~\cite{susarla2022hyperspectral}.

In the bilayers investigated here, we did not observe any optical spectroscopic signatures of correlated insulating states in hole-doped 57.5$^{\circ}$ MoSe$_2$ bilayers, often observed in W-based bilayers \cite{miao2021strong,wang2023intercell}. To rule out sample quality as a possible reason, we have performed combined optical spectroscopy and microwave impedance microscopy (MIM) measurements on another sample (Device 4, data shown in Fig.~S10). MIM has previously been established as a sensitive probe of correlated electronic states in WSe$_2$/WS$_2$ moir\'e bilayers~\cite{chu2020nanoscale,huang2021correlated}. No insulating states in the 57.5$^{\circ}$ MoSe$_2$ bilayers were observed using either of these two techniques.

In conclusion, our experiments provide spectroscopic evidence for the complexity of dopants and exciton distribution in both momentum and real space in moir\'e superlattices. The emergence of the novel charge transfer trion arises from the flat valence band and the distinct moir\'e potential experienced by doped holes compared to optically generated electron-hole pairs in twisted MoSe$_2$ bilayers. Future experiments with higher spatial resolution (e.g. scanning tunneling microscopy) or direct access to momentum space (time-resolved angle-resolved photoemission spectroscopy) are required to deliver more visually compelling confirmation.
Unlike interlayer moir\'e trions~\cite{wang2023intercell,wang2021moire,liu2021signatures,brotons2021moire}, the new charge transfer trions feature strong oscillator strength and complex quantum indices including spin, momentum, and valley.  These findings may open new pathways toward coherent control of localized spin arrays for quantum simulation~\cite{xiong2023correlated} and quantum information processing analogous to those in conventional quantum dots~\cite{kroutvar2004optically}.

\section*{Data availability}
Source data are provided with this paper~\cite{liu2026figshare}. The data that support the plots within this paper and other findings of this study are available from the corresponding authors upon reasonable requests.\\ 

\section*{Supplementary Material}
Supplementary material contains the details of experimental techniques and Extended Figures 1-13 as well as references~\cite{regan2020mott,lim2023modulation,shimazaki2020strongly,giannozzi2009quantum,PhysRevB.34.5390,PhysRevB.62.4927,van2019probing,thompson2022lammps,jiang2017handbook,naik2019kolmogorov,PhysRevB.96.235441,geick1966normal,zhang2014absorption,born2013principles}.

\section*{Acknowledgments}
The experiments were primarily supported by the National Science Foundation through the MRSEC program under Cooperative Agreement No. DMR-2308817, which enabled collaboration among Shih, Warner, MacDonald, Baldini, and Li and supported the facility where some measurements were performed. X. Li acknowledges support from the Office of Naval Research (grant N000142512069) and the Welch Foundation Chair F-0014 for sample preparation. X. Liu, D. Huang, and Y. Ni were partially supported by the Department of Energy, Office of Basic Energy Sciences (grant DE-SC0019398), and X.C. Liu by the NSF EPM program (DMR-2225645) for device fabrication. X. Yu's work on field-dependent measurements is primarily supported by the Army Research Office via grant W911NF-25-1-0058. The collaboration between UT Austin (Li and Lai) and Washington University was supported by the NSF Designing Materials to Revolutionize and Engineer our Future (DMREF) program under grants DMR-2118806 and DMR-2118779. H.W. was supported by NSF grant DMR-2124934. Work in the Baldini group at UT Austin was primarily supported by the Robert A. Welch Foundation under grant F-2092-20250403 (to F.Y.G. for data taking and analysis) and the United States Army Research Office (W911NF-23-1-0394) (to E.B. for experimental support and manuscript editing). K.L. and H.Y. acknowledge support from the United States Army Research Office under Grant No. W911NF-25-1-0232. K.W. and T.T. acknowledge support from the JSPS KAKENHI (Grant Numbers 20H00354, 21H05233 and 23H02052) and World Premier International Research Center Initiative (WPI), MEXT, Japan. The computing resource is provided by Purdue Anvil at the Rosen Center for Advanced Computing through allocation DMR100005 from the Advanced Cyberinfrastructure Coordination Ecosystem: Services Support (ACCESS) program, which is supported by National Science Foundation grants 2138259, 2138286, 2138307, 2137603, and 2138296.

\onecolumngrid


\begin{thebibliography}{64}%
\makeatletter
\providecommand \@ifxundefined [1]{%
 \@ifx{#1\undefined}
}%
\providecommand \@ifnum [1]{%
 \ifnum #1\expandafter \@firstoftwo
 \else \expandafter \@secondoftwo
 \fi
}%
\providecommand \@ifx [1]{%
 \ifx #1\expandafter \@firstoftwo
 \else \expandafter \@secondoftwo
 \fi
}%
\providecommand \natexlab [1]{#1}%
\providecommand \enquote  [1]{``#1''}%
\providecommand \bibnamefont  [1]{#1}%
\providecommand \bibfnamefont [1]{#1}%
\providecommand \citenamefont [1]{#1}%
\providecommand \href@noop [0]{\@secondoftwo}%
\providecommand \href [0]{\begingroup \@sanitize@url \@href}%
\providecommand \@href[1]{\@@startlink{#1}\@@href}%
\providecommand \@@href[1]{\endgroup#1\@@endlink}%
\providecommand \@sanitize@url [0]{\catcode `\\12\catcode `\$12\catcode `\&12\catcode `\#12\catcode `\^12\catcode `\_12\catcode `\%12\relax}%
\providecommand \@@startlink[1]{}%
\providecommand \@@endlink[0]{}%
\providecommand \url  [0]{\begingroup\@sanitize@url \@url }%
\providecommand \@url [1]{\endgroup\@href {#1}{\urlprefix }}%
\providecommand \urlprefix  [0]{URL }%
\providecommand \Eprint [0]{\href }%
\providecommand \doibase [0]{https://doi.org/}%
\providecommand \selectlanguage [0]{\@gobble}%
\providecommand \bibinfo  [0]{\@secondoftwo}%
\providecommand \bibfield  [0]{\@secondoftwo}%
\providecommand \translation [1]{[#1]}%
\providecommand \BibitemOpen [0]{}%
\providecommand \bibitemStop [0]{}%
\providecommand \bibitemNoStop [0]{.\EOS\space}%
\providecommand \EOS [0]{\spacefactor3000\relax}%
\providecommand \BibitemShut  [1]{\csname bibitem#1\endcsname}%
\let\auto@bib@innerbib\@empty
\bibitem [{\citenamefont {Wang}\ \emph {et~al.}(2020)\citenamefont {Wang}, \citenamefont {Shih}, \citenamefont {Ghiotto}, \citenamefont {Xian}, \citenamefont {Rhodes}, \citenamefont {Tan}, \citenamefont {Claassen}, \citenamefont {Kennes}, \citenamefont {Bai}, \citenamefont {Kim} \emph {et~al.}}]{wang2020correlated}%
  \BibitemOpen
  \bibfield  {author} {\bibinfo {author} {\bibfnamefont {L.}~\bibnamefont {Wang}}, \bibinfo {author} {\bibfnamefont {E.-M.}\ \bibnamefont {Shih}}, \bibinfo {author} {\bibfnamefont {A.}~\bibnamefont {Ghiotto}}, \bibinfo {author} {\bibfnamefont {L.}~\bibnamefont {Xian}}, \bibinfo {author} {\bibfnamefont {D.~A.}\ \bibnamefont {Rhodes}}, \bibinfo {author} {\bibfnamefont {C.}~\bibnamefont {Tan}}, \bibinfo {author} {\bibfnamefont {M.}~\bibnamefont {Claassen}}, \bibinfo {author} {\bibfnamefont {D.~M.}\ \bibnamefont {Kennes}}, \bibinfo {author} {\bibfnamefont {Y.}~\bibnamefont {Bai}}, \bibinfo {author} {\bibfnamefont {B.}~\bibnamefont {Kim}}, \emph {et~al.},\ }\href@noop {} {\bibfield  {journal} {\bibinfo  {journal} {Nature Materials}\ }\textbf {\bibinfo {volume} {19}},\ \bibinfo {pages} {861} (\bibinfo {year} {2020})}\BibitemShut {NoStop}%
\bibitem [{\citenamefont {Tang}\ \emph {et~al.}(2020)\citenamefont {Tang}, \citenamefont {Li}, \citenamefont {Li}, \citenamefont {Xu}, \citenamefont {Liu}, \citenamefont {Barmak}, \citenamefont {Watanabe}, \citenamefont {Taniguchi}, \citenamefont {MacDonald}, \citenamefont {Shan} \emph {et~al.}}]{tang2020simulation}%
  \BibitemOpen
  \bibfield  {author} {\bibinfo {author} {\bibfnamefont {Y.}~\bibnamefont {Tang}}, \bibinfo {author} {\bibfnamefont {L.}~\bibnamefont {Li}}, \bibinfo {author} {\bibfnamefont {T.}~\bibnamefont {Li}}, \bibinfo {author} {\bibfnamefont {Y.}~\bibnamefont {Xu}}, \bibinfo {author} {\bibfnamefont {S.}~\bibnamefont {Liu}}, \bibinfo {author} {\bibfnamefont {K.}~\bibnamefont {Barmak}}, \bibinfo {author} {\bibfnamefont {K.}~\bibnamefont {Watanabe}}, \bibinfo {author} {\bibfnamefont {T.}~\bibnamefont {Taniguchi}}, \bibinfo {author} {\bibfnamefont {A.~H.}\ \bibnamefont {MacDonald}}, \bibinfo {author} {\bibfnamefont {J.}~\bibnamefont {Shan}}, \emph {et~al.},\ }\href@noop {} {\bibfield  {journal} {\bibinfo  {journal} {Nature}\ }\textbf {\bibinfo {volume} {579}},\ \bibinfo {pages} {353} (\bibinfo {year} {2020})}\BibitemShut {NoStop}%
\bibitem [{\citenamefont {Shimazaki}\ \emph {et~al.}(2020)\citenamefont {Shimazaki}, \citenamefont {Schwartz}, \citenamefont {Watanabe}, \citenamefont {Taniguchi}, \citenamefont {Kroner},\ and\ \citenamefont {Imamo{\u{g}}lu}}]{shimazaki2020strongly}%
  \BibitemOpen
  \bibfield  {author} {\bibinfo {author} {\bibfnamefont {Y.}~\bibnamefont {Shimazaki}}, \bibinfo {author} {\bibfnamefont {I.}~\bibnamefont {Schwartz}}, \bibinfo {author} {\bibfnamefont {K.}~\bibnamefont {Watanabe}}, \bibinfo {author} {\bibfnamefont {T.}~\bibnamefont {Taniguchi}}, \bibinfo {author} {\bibfnamefont {M.}~\bibnamefont {Kroner}},\ and\ \bibinfo {author} {\bibfnamefont {A.}~\bibnamefont {Imamo{\u{g}}lu}},\ }\href@noop {} {\bibfield  {journal} {\bibinfo  {journal} {Nature}\ }\textbf {\bibinfo {volume} {580}},\ \bibinfo {pages} {472} (\bibinfo {year} {2020})}\BibitemShut {NoStop}%
\bibitem [{\citenamefont {Regan}\ \emph {et~al.}(2020)\citenamefont {Regan}, \citenamefont {Wang}, \citenamefont {Jin}, \citenamefont {Bakti~Utama}, \citenamefont {Gao}, \citenamefont {Wei}, \citenamefont {Zhao}, \citenamefont {Zhao}, \citenamefont {Zhang}, \citenamefont {Yumigeta} \emph {et~al.}}]{regan2020mott}%
  \BibitemOpen
  \bibfield  {author} {\bibinfo {author} {\bibfnamefont {E.~C.}\ \bibnamefont {Regan}}, \bibinfo {author} {\bibfnamefont {D.}~\bibnamefont {Wang}}, \bibinfo {author} {\bibfnamefont {C.}~\bibnamefont {Jin}}, \bibinfo {author} {\bibfnamefont {M.~I.}\ \bibnamefont {Bakti~Utama}}, \bibinfo {author} {\bibfnamefont {B.}~\bibnamefont {Gao}}, \bibinfo {author} {\bibfnamefont {X.}~\bibnamefont {Wei}}, \bibinfo {author} {\bibfnamefont {S.}~\bibnamefont {Zhao}}, \bibinfo {author} {\bibfnamefont {W.}~\bibnamefont {Zhao}}, \bibinfo {author} {\bibfnamefont {Z.}~\bibnamefont {Zhang}}, \bibinfo {author} {\bibfnamefont {K.}~\bibnamefont {Yumigeta}}, \emph {et~al.},\ }\href@noop {} {\bibfield  {journal} {\bibinfo  {journal} {Nature}\ }\textbf {\bibinfo {volume} {579}},\ \bibinfo {pages} {359} (\bibinfo {year} {2020})}\BibitemShut {NoStop}%
\bibitem [{\citenamefont {Regan}\ \emph {et~al.}(2022)\citenamefont {Regan}, \citenamefont {Wang}, \citenamefont {Paik}, \citenamefont {Zeng}, \citenamefont {Zhang}, \citenamefont {Zhu}, \citenamefont {MacDonald}, \citenamefont {Deng},\ and\ \citenamefont {Wang}}]{regan2022emerging}%
  \BibitemOpen
  \bibfield  {author} {\bibinfo {author} {\bibfnamefont {E.~C.}\ \bibnamefont {Regan}}, \bibinfo {author} {\bibfnamefont {D.}~\bibnamefont {Wang}}, \bibinfo {author} {\bibfnamefont {E.~Y.}\ \bibnamefont {Paik}}, \bibinfo {author} {\bibfnamefont {Y.}~\bibnamefont {Zeng}}, \bibinfo {author} {\bibfnamefont {L.}~\bibnamefont {Zhang}}, \bibinfo {author} {\bibfnamefont {J.}~\bibnamefont {Zhu}}, \bibinfo {author} {\bibfnamefont {A.~H.}\ \bibnamefont {MacDonald}}, \bibinfo {author} {\bibfnamefont {H.}~\bibnamefont {Deng}},\ and\ \bibinfo {author} {\bibfnamefont {F.}~\bibnamefont {Wang}},\ }\href@noop {} {\bibfield  {journal} {\bibinfo  {journal} {Nature Reviews Materials}\ }\textbf {\bibinfo {volume} {7}},\ \bibinfo {pages} {778} (\bibinfo {year} {2022})}\BibitemShut {NoStop}%
\bibitem [{\citenamefont {Huang}\ \emph {et~al.}(2022)\citenamefont {Huang}, \citenamefont {Choi}, \citenamefont {Shih},\ and\ \citenamefont {Li}}]{huang2022excitons}%
  \BibitemOpen
  \bibfield  {author} {\bibinfo {author} {\bibfnamefont {D.}~\bibnamefont {Huang}}, \bibinfo {author} {\bibfnamefont {J.}~\bibnamefont {Choi}}, \bibinfo {author} {\bibfnamefont {C.-K.}\ \bibnamefont {Shih}},\ and\ \bibinfo {author} {\bibfnamefont {X.}~\bibnamefont {Li}},\ }\href@noop {} {\bibfield  {journal} {\bibinfo  {journal} {Nature Nanotechnology}\ }\textbf {\bibinfo {volume} {17}},\ \bibinfo {pages} {227} (\bibinfo {year} {2022})}\BibitemShut {NoStop}%
\bibitem [{\citenamefont {Andersen}\ \emph {et~al.}(2021)\citenamefont {Andersen}, \citenamefont {Scuri}, \citenamefont {Sushko}, \citenamefont {De~Greve}, \citenamefont {Sung}, \citenamefont {Zhou}, \citenamefont {Wild}, \citenamefont {Gelly}, \citenamefont {Heo}, \citenamefont {B{\'e}rub{\'e}} \emph {et~al.}}]{andersen2021excitons}%
  \BibitemOpen
  \bibfield  {author} {\bibinfo {author} {\bibfnamefont {T.~I.}\ \bibnamefont {Andersen}}, \bibinfo {author} {\bibfnamefont {G.}~\bibnamefont {Scuri}}, \bibinfo {author} {\bibfnamefont {A.}~\bibnamefont {Sushko}}, \bibinfo {author} {\bibfnamefont {K.}~\bibnamefont {De~Greve}}, \bibinfo {author} {\bibfnamefont {J.}~\bibnamefont {Sung}}, \bibinfo {author} {\bibfnamefont {Y.}~\bibnamefont {Zhou}}, \bibinfo {author} {\bibfnamefont {D.~S.}\ \bibnamefont {Wild}}, \bibinfo {author} {\bibfnamefont {R.~J.}\ \bibnamefont {Gelly}}, \bibinfo {author} {\bibfnamefont {H.}~\bibnamefont {Heo}}, \bibinfo {author} {\bibfnamefont {D.}~\bibnamefont {B{\'e}rub{\'e}}}, \emph {et~al.},\ }\href@noop {} {\bibfield  {journal} {\bibinfo  {journal} {Nature Materials}\ }\textbf {\bibinfo {volume} {20}},\ \bibinfo {pages} {480} (\bibinfo {year} {2021})}\BibitemShut {NoStop}%
\bibitem [{\citenamefont {Tran}\ \emph {et~al.}(2019)\citenamefont {Tran}, \citenamefont {Moody}, \citenamefont {Wu}, \citenamefont {Lu}, \citenamefont {Choi}, \citenamefont {Kim}, \citenamefont {Rai}, \citenamefont {Sanchez}, \citenamefont {Quan}, \citenamefont {Singh} \emph {et~al.}}]{tran2019evidence}%
  \BibitemOpen
  \bibfield  {author} {\bibinfo {author} {\bibfnamefont {K.}~\bibnamefont {Tran}}, \bibinfo {author} {\bibfnamefont {G.}~\bibnamefont {Moody}}, \bibinfo {author} {\bibfnamefont {F.}~\bibnamefont {Wu}}, \bibinfo {author} {\bibfnamefont {X.}~\bibnamefont {Lu}}, \bibinfo {author} {\bibfnamefont {J.}~\bibnamefont {Choi}}, \bibinfo {author} {\bibfnamefont {K.}~\bibnamefont {Kim}}, \bibinfo {author} {\bibfnamefont {A.}~\bibnamefont {Rai}}, \bibinfo {author} {\bibfnamefont {D.~A.}\ \bibnamefont {Sanchez}}, \bibinfo {author} {\bibfnamefont {J.}~\bibnamefont {Quan}}, \bibinfo {author} {\bibfnamefont {A.}~\bibnamefont {Singh}}, \emph {et~al.},\ }\href@noop {} {\bibfield  {journal} {\bibinfo  {journal} {Nature}\ }\textbf {\bibinfo {volume} {567}},\ \bibinfo {pages} {71} (\bibinfo {year} {2019})}\BibitemShut {NoStop}%
\bibitem [{\citenamefont {Seyler}\ \emph {et~al.}(2019)\citenamefont {Seyler}, \citenamefont {Rivera}, \citenamefont {Yu}, \citenamefont {Wilson}, \citenamefont {Ray}, \citenamefont {Mandrus}, \citenamefont {Yan}, \citenamefont {Yao},\ and\ \citenamefont {Xu}}]{seyler2019signatures}%
  \BibitemOpen
  \bibfield  {author} {\bibinfo {author} {\bibfnamefont {K.~L.}\ \bibnamefont {Seyler}}, \bibinfo {author} {\bibfnamefont {P.}~\bibnamefont {Rivera}}, \bibinfo {author} {\bibfnamefont {H.}~\bibnamefont {Yu}}, \bibinfo {author} {\bibfnamefont {N.~P.}\ \bibnamefont {Wilson}}, \bibinfo {author} {\bibfnamefont {E.~L.}\ \bibnamefont {Ray}}, \bibinfo {author} {\bibfnamefont {D.~G.}\ \bibnamefont {Mandrus}}, \bibinfo {author} {\bibfnamefont {J.}~\bibnamefont {Yan}}, \bibinfo {author} {\bibfnamefont {W.}~\bibnamefont {Yao}},\ and\ \bibinfo {author} {\bibfnamefont {X.}~\bibnamefont {Xu}},\ }\href@noop {} {\bibfield  {journal} {\bibinfo  {journal} {Nature}\ }\textbf {\bibinfo {volume} {567}},\ \bibinfo {pages} {66} (\bibinfo {year} {2019})}\BibitemShut {NoStop}%
\bibitem [{\citenamefont {Jin}\ \emph {et~al.}(2019)\citenamefont {Jin}, \citenamefont {Regan}, \citenamefont {Yan}, \citenamefont {Iqbal Bakti~Utama}, \citenamefont {Wang}, \citenamefont {Zhao}, \citenamefont {Qin}, \citenamefont {Yang}, \citenamefont {Zheng}, \citenamefont {Shi} \emph {et~al.}}]{jin2019observation}%
  \BibitemOpen
  \bibfield  {author} {\bibinfo {author} {\bibfnamefont {C.}~\bibnamefont {Jin}}, \bibinfo {author} {\bibfnamefont {E.~C.}\ \bibnamefont {Regan}}, \bibinfo {author} {\bibfnamefont {A.}~\bibnamefont {Yan}}, \bibinfo {author} {\bibfnamefont {M.}~\bibnamefont {Iqbal Bakti~Utama}}, \bibinfo {author} {\bibfnamefont {D.}~\bibnamefont {Wang}}, \bibinfo {author} {\bibfnamefont {S.}~\bibnamefont {Zhao}}, \bibinfo {author} {\bibfnamefont {Y.}~\bibnamefont {Qin}}, \bibinfo {author} {\bibfnamefont {S.}~\bibnamefont {Yang}}, \bibinfo {author} {\bibfnamefont {Z.}~\bibnamefont {Zheng}}, \bibinfo {author} {\bibfnamefont {S.}~\bibnamefont {Shi}}, \emph {et~al.},\ }\href@noop {} {\bibfield  {journal} {\bibinfo  {journal} {Nature}\ }\textbf {\bibinfo {volume} {567}},\ \bibinfo {pages} {76} (\bibinfo {year} {2019})}\BibitemShut {NoStop}%
\bibitem [{\citenamefont {Alexeev}\ \emph {et~al.}(2019)\citenamefont {Alexeev}, \citenamefont {Ruiz-Tijerina}, \citenamefont {Danovich}, \citenamefont {Hamer}, \citenamefont {Terry}, \citenamefont {Nayak}, \citenamefont {Ahn}, \citenamefont {Pak}, \citenamefont {Lee}, \citenamefont {Sohn} \emph {et~al.}}]{alexeev2019resonantly}%
  \BibitemOpen
  \bibfield  {author} {\bibinfo {author} {\bibfnamefont {E.~M.}\ \bibnamefont {Alexeev}}, \bibinfo {author} {\bibfnamefont {D.~A.}\ \bibnamefont {Ruiz-Tijerina}}, \bibinfo {author} {\bibfnamefont {M.}~\bibnamefont {Danovich}}, \bibinfo {author} {\bibfnamefont {M.~J.}\ \bibnamefont {Hamer}}, \bibinfo {author} {\bibfnamefont {D.~J.}\ \bibnamefont {Terry}}, \bibinfo {author} {\bibfnamefont {P.~K.}\ \bibnamefont {Nayak}}, \bibinfo {author} {\bibfnamefont {S.}~\bibnamefont {Ahn}}, \bibinfo {author} {\bibfnamefont {S.}~\bibnamefont {Pak}}, \bibinfo {author} {\bibfnamefont {J.}~\bibnamefont {Lee}}, \bibinfo {author} {\bibfnamefont {J.~I.}\ \bibnamefont {Sohn}}, \emph {et~al.},\ }\href@noop {} {\bibfield  {journal} {\bibinfo  {journal} {Nature}\ }\textbf {\bibinfo {volume} {567}},\ \bibinfo {pages} {81} (\bibinfo {year} {2019})}\BibitemShut {NoStop}%
\bibitem [{\citenamefont {Zhang}\ \emph {et~al.}(2018)\citenamefont {Zhang}, \citenamefont {Surrente}, \citenamefont {Baranowski}, \citenamefont {Maude}, \citenamefont {Gant}, \citenamefont {Castellanos-Gomez},\ and\ \citenamefont {Plochocka}}]{zhang2018moire}%
  \BibitemOpen
  \bibfield  {author} {\bibinfo {author} {\bibfnamefont {N.}~\bibnamefont {Zhang}}, \bibinfo {author} {\bibfnamefont {A.}~\bibnamefont {Surrente}}, \bibinfo {author} {\bibfnamefont {M.}~\bibnamefont {Baranowski}}, \bibinfo {author} {\bibfnamefont {D.~K.}\ \bibnamefont {Maude}}, \bibinfo {author} {\bibfnamefont {P.}~\bibnamefont {Gant}}, \bibinfo {author} {\bibfnamefont {A.}~\bibnamefont {Castellanos-Gomez}},\ and\ \bibinfo {author} {\bibfnamefont {P.}~\bibnamefont {Plochocka}},\ }\href@noop {} {\bibfield  {journal} {\bibinfo  {journal} {Nano letters}\ }\textbf {\bibinfo {volume} {18}},\ \bibinfo {pages} {7651} (\bibinfo {year} {2018})}\BibitemShut {NoStop}%
\bibitem [{\citenamefont {Merkl}\ \emph {et~al.}(2020)\citenamefont {Merkl}, \citenamefont {Mooshammer}, \citenamefont {Brem}, \citenamefont {Girnghuber}, \citenamefont {Lin}, \citenamefont {Weigl}, \citenamefont {Liebich}, \citenamefont {Yong}, \citenamefont {Gillen}, \citenamefont {Maultzsch} \emph {et~al.}}]{merkl2020twist}%
  \BibitemOpen
  \bibfield  {author} {\bibinfo {author} {\bibfnamefont {P.}~\bibnamefont {Merkl}}, \bibinfo {author} {\bibfnamefont {F.}~\bibnamefont {Mooshammer}}, \bibinfo {author} {\bibfnamefont {S.}~\bibnamefont {Brem}}, \bibinfo {author} {\bibfnamefont {A.}~\bibnamefont {Girnghuber}}, \bibinfo {author} {\bibfnamefont {K.-Q.}\ \bibnamefont {Lin}}, \bibinfo {author} {\bibfnamefont {L.}~\bibnamefont {Weigl}}, \bibinfo {author} {\bibfnamefont {M.}~\bibnamefont {Liebich}}, \bibinfo {author} {\bibfnamefont {C.-K.}\ \bibnamefont {Yong}}, \bibinfo {author} {\bibfnamefont {R.}~\bibnamefont {Gillen}}, \bibinfo {author} {\bibfnamefont {J.}~\bibnamefont {Maultzsch}}, \emph {et~al.},\ }\href@noop {} {\bibfield  {journal} {\bibinfo  {journal} {Nature Communications}\ }\textbf {\bibinfo {volume} {11}},\ \bibinfo {pages} {2167} (\bibinfo {year} {2020})}\BibitemShut {NoStop}%
\bibitem [{\citenamefont {Choi}\ \emph {et~al.}(2021)\citenamefont {Choi}, \citenamefont {Florian}, \citenamefont {Steinhoff}, \citenamefont {Erben}, \citenamefont {Tran}, \citenamefont {Kim}, \citenamefont {Sun}, \citenamefont {Quan}, \citenamefont {Claassen}, \citenamefont {Majumder} \emph {et~al.}}]{choi2021twist}%
  \BibitemOpen
  \bibfield  {author} {\bibinfo {author} {\bibfnamefont {J.}~\bibnamefont {Choi}}, \bibinfo {author} {\bibfnamefont {M.}~\bibnamefont {Florian}}, \bibinfo {author} {\bibfnamefont {A.}~\bibnamefont {Steinhoff}}, \bibinfo {author} {\bibfnamefont {D.}~\bibnamefont {Erben}}, \bibinfo {author} {\bibfnamefont {K.}~\bibnamefont {Tran}}, \bibinfo {author} {\bibfnamefont {D.~S.}\ \bibnamefont {Kim}}, \bibinfo {author} {\bibfnamefont {L.}~\bibnamefont {Sun}}, \bibinfo {author} {\bibfnamefont {J.}~\bibnamefont {Quan}}, \bibinfo {author} {\bibfnamefont {R.}~\bibnamefont {Claassen}}, \bibinfo {author} {\bibfnamefont {S.}~\bibnamefont {Majumder}}, \emph {et~al.},\ }\href@noop {} {\bibfield  {journal} {\bibinfo  {journal} {Physical Review Letters}\ }\textbf {\bibinfo {volume} {126}},\ \bibinfo {pages} {047401} (\bibinfo {year} {2021})}\BibitemShut {NoStop}%
\bibitem [{\citenamefont {Zhang}\ \emph {et~al.}(2020{\natexlab{a}})\citenamefont {Zhang}, \citenamefont {Zhang}, \citenamefont {Wu}, \citenamefont {Wang}, \citenamefont {Gogna}, \citenamefont {Hou}, \citenamefont {Watanabe}, \citenamefont {Taniguchi}, \citenamefont {Kulkarni}, \citenamefont {Kuo} \emph {et~al.}}]{zhang2020twist}%
  \BibitemOpen
  \bibfield  {author} {\bibinfo {author} {\bibfnamefont {L.}~\bibnamefont {Zhang}}, \bibinfo {author} {\bibfnamefont {Z.}~\bibnamefont {Zhang}}, \bibinfo {author} {\bibfnamefont {F.}~\bibnamefont {Wu}}, \bibinfo {author} {\bibfnamefont {D.}~\bibnamefont {Wang}}, \bibinfo {author} {\bibfnamefont {R.}~\bibnamefont {Gogna}}, \bibinfo {author} {\bibfnamefont {S.}~\bibnamefont {Hou}}, \bibinfo {author} {\bibfnamefont {K.}~\bibnamefont {Watanabe}}, \bibinfo {author} {\bibfnamefont {T.}~\bibnamefont {Taniguchi}}, \bibinfo {author} {\bibfnamefont {K.}~\bibnamefont {Kulkarni}}, \bibinfo {author} {\bibfnamefont {T.}~\bibnamefont {Kuo}}, \emph {et~al.},\ }\href@noop {} {\bibfield  {journal} {\bibinfo  {journal} {Nature Communications}\ }\textbf {\bibinfo {volume} {11}},\ \bibinfo {pages} {5888} (\bibinfo {year} {2020}{\natexlab{a}})}\BibitemShut {NoStop}%
\bibitem [{\citenamefont {Lin}\ \emph {et~al.}(2024)\citenamefont {Lin}, \citenamefont {Fang}, \citenamefont {Kalaboukhov}, \citenamefont {Liu}, \citenamefont {Zhang}, \citenamefont {Fischer}, \citenamefont {Li}, \citenamefont {Hagel}, \citenamefont {Brem}, \citenamefont {Malic} \emph {et~al.}}]{lin2024moire}%
  \BibitemOpen
  \bibfield  {author} {\bibinfo {author} {\bibfnamefont {Q.}~\bibnamefont {Lin}}, \bibinfo {author} {\bibfnamefont {H.}~\bibnamefont {Fang}}, \bibinfo {author} {\bibfnamefont {A.}~\bibnamefont {Kalaboukhov}}, \bibinfo {author} {\bibfnamefont {Y.}~\bibnamefont {Liu}}, \bibinfo {author} {\bibfnamefont {Y.}~\bibnamefont {Zhang}}, \bibinfo {author} {\bibfnamefont {M.}~\bibnamefont {Fischer}}, \bibinfo {author} {\bibfnamefont {J.}~\bibnamefont {Li}}, \bibinfo {author} {\bibfnamefont {J.}~\bibnamefont {Hagel}}, \bibinfo {author} {\bibfnamefont {S.}~\bibnamefont {Brem}}, \bibinfo {author} {\bibfnamefont {E.}~\bibnamefont {Malic}}, \emph {et~al.},\ }\href@noop {} {\bibfield  {journal} {\bibinfo  {journal} {Nature Communications}\ }\textbf {\bibinfo {volume} {15}},\ \bibinfo {pages} {8762} (\bibinfo {year} {2024})}\BibitemShut {NoStop}%
\bibitem [{\citenamefont {Xiong}\ \emph {et~al.}(2023)\citenamefont {Xiong}, \citenamefont {Nie}, \citenamefont {Brantly}, \citenamefont {Hays}, \citenamefont {Sailus}, \citenamefont {Watanabe}, \citenamefont {Taniguchi}, \citenamefont {Tongay},\ and\ \citenamefont {Jin}}]{xiong2023correlated}%
  \BibitemOpen
  \bibfield  {author} {\bibinfo {author} {\bibfnamefont {R.}~\bibnamefont {Xiong}}, \bibinfo {author} {\bibfnamefont {J.~H.}\ \bibnamefont {Nie}}, \bibinfo {author} {\bibfnamefont {S.~L.}\ \bibnamefont {Brantly}}, \bibinfo {author} {\bibfnamefont {P.}~\bibnamefont {Hays}}, \bibinfo {author} {\bibfnamefont {R.}~\bibnamefont {Sailus}}, \bibinfo {author} {\bibfnamefont {K.}~\bibnamefont {Watanabe}}, \bibinfo {author} {\bibfnamefont {T.}~\bibnamefont {Taniguchi}}, \bibinfo {author} {\bibfnamefont {S.}~\bibnamefont {Tongay}},\ and\ \bibinfo {author} {\bibfnamefont {C.}~\bibnamefont {Jin}},\ }\href@noop {} {\bibfield  {journal} {\bibinfo  {journal} {Science}\ ,\ \bibinfo {pages} {eadd5574}} (\bibinfo {year} {2023})}\BibitemShut {NoStop}%
\bibitem [{\citenamefont {Xu}\ \emph {et~al.}(2020)\citenamefont {Xu}, \citenamefont {Liu}, \citenamefont {Rhodes}, \citenamefont {Watanabe}, \citenamefont {Taniguchi}, \citenamefont {Hone}, \citenamefont {Elser}, \citenamefont {Mak},\ and\ \citenamefont {Shan}}]{xu2020correlated}%
  \BibitemOpen
  \bibfield  {author} {\bibinfo {author} {\bibfnamefont {Y.}~\bibnamefont {Xu}}, \bibinfo {author} {\bibfnamefont {S.}~\bibnamefont {Liu}}, \bibinfo {author} {\bibfnamefont {D.~A.}\ \bibnamefont {Rhodes}}, \bibinfo {author} {\bibfnamefont {K.}~\bibnamefont {Watanabe}}, \bibinfo {author} {\bibfnamefont {T.}~\bibnamefont {Taniguchi}}, \bibinfo {author} {\bibfnamefont {J.}~\bibnamefont {Hone}}, \bibinfo {author} {\bibfnamefont {V.}~\bibnamefont {Elser}}, \bibinfo {author} {\bibfnamefont {K.~F.}\ \bibnamefont {Mak}},\ and\ \bibinfo {author} {\bibfnamefont {J.}~\bibnamefont {Shan}},\ }\href@noop {} {\bibfield  {journal} {\bibinfo  {journal} {Nature}\ }\textbf {\bibinfo {volume} {587}},\ \bibinfo {pages} {214} (\bibinfo {year} {2020})}\BibitemShut {NoStop}%
\bibitem [{\citenamefont {Wu}\ \emph {et~al.}(2017)\citenamefont {Wu}, \citenamefont {Lovorn},\ and\ \citenamefont {MacDonald}}]{wu2017topological}%
  \BibitemOpen
  \bibfield  {author} {\bibinfo {author} {\bibfnamefont {F.}~\bibnamefont {Wu}}, \bibinfo {author} {\bibfnamefont {T.}~\bibnamefont {Lovorn}},\ and\ \bibinfo {author} {\bibfnamefont {A.~H.}\ \bibnamefont {MacDonald}},\ }\href@noop {} {\bibfield  {journal} {\bibinfo  {journal} {Physical Review Letters}\ }\textbf {\bibinfo {volume} {118}},\ \bibinfo {pages} {147401} (\bibinfo {year} {2017})}\BibitemShut {NoStop}%
\bibitem [{\citenamefont {Brem}\ \emph {et~al.}(2020)\citenamefont {Brem}, \citenamefont {Linderalv}, \citenamefont {Erhart},\ and\ \citenamefont {Malic}}]{brem2020tunable}%
  \BibitemOpen
  \bibfield  {author} {\bibinfo {author} {\bibfnamefont {S.}~\bibnamefont {Brem}}, \bibinfo {author} {\bibfnamefont {C.}~\bibnamefont {Linderalv}}, \bibinfo {author} {\bibfnamefont {P.}~\bibnamefont {Erhart}},\ and\ \bibinfo {author} {\bibfnamefont {E.}~\bibnamefont {Malic}},\ }\href@noop {} {\bibfield  {journal} {\bibinfo  {journal} {Nano Letters}\ }\textbf {\bibinfo {volume} {20}},\ \bibinfo {pages} {8534} (\bibinfo {year} {2020})}\BibitemShut {NoStop}%
\bibitem [{\citenamefont {Zhang}\ \emph {et~al.}(2020{\natexlab{b}})\citenamefont {Zhang}, \citenamefont {Wang}, \citenamefont {Watanabe}, \citenamefont {Taniguchi}, \citenamefont {Ueno}, \citenamefont {Tutuc},\ and\ \citenamefont {LeRoy}}]{zhang2020flat}%
  \BibitemOpen
  \bibfield  {author} {\bibinfo {author} {\bibfnamefont {Z.}~\bibnamefont {Zhang}}, \bibinfo {author} {\bibfnamefont {Y.}~\bibnamefont {Wang}}, \bibinfo {author} {\bibfnamefont {K.}~\bibnamefont {Watanabe}}, \bibinfo {author} {\bibfnamefont {T.}~\bibnamefont {Taniguchi}}, \bibinfo {author} {\bibfnamefont {K.}~\bibnamefont {Ueno}}, \bibinfo {author} {\bibfnamefont {E.}~\bibnamefont {Tutuc}},\ and\ \bibinfo {author} {\bibfnamefont {B.~J.}\ \bibnamefont {LeRoy}},\ }\href@noop {} {\bibfield  {journal} {\bibinfo  {journal} {Nature Physics}\ }\textbf {\bibinfo {volume} {16}},\ \bibinfo {pages} {1093} (\bibinfo {year} {2020}{\natexlab{b}})}\BibitemShut {NoStop}%
\bibitem [{\citenamefont {Li}\ \emph {et~al.}(2021)\citenamefont {Li}, \citenamefont {Li}, \citenamefont {Naik}, \citenamefont {Xie}, \citenamefont {Li}, \citenamefont {Regan}, \citenamefont {Wang}, \citenamefont {Zhao}, \citenamefont {Yumigeta}, \citenamefont {Blei} \emph {et~al.}}]{li2021imagingdischarge}%
  \BibitemOpen
  \bibfield  {author} {\bibinfo {author} {\bibfnamefont {H.}~\bibnamefont {Li}}, \bibinfo {author} {\bibfnamefont {S.}~\bibnamefont {Li}}, \bibinfo {author} {\bibfnamefont {M.~H.}\ \bibnamefont {Naik}}, \bibinfo {author} {\bibfnamefont {J.}~\bibnamefont {Xie}}, \bibinfo {author} {\bibfnamefont {X.}~\bibnamefont {Li}}, \bibinfo {author} {\bibfnamefont {E.}~\bibnamefont {Regan}}, \bibinfo {author} {\bibfnamefont {D.}~\bibnamefont {Wang}}, \bibinfo {author} {\bibfnamefont {W.}~\bibnamefont {Zhao}}, \bibinfo {author} {\bibfnamefont {K.}~\bibnamefont {Yumigeta}}, \bibinfo {author} {\bibfnamefont {M.}~\bibnamefont {Blei}}, \emph {et~al.},\ }\href@noop {} {\bibfield  {journal} {\bibinfo  {journal} {Nature Physics}\ }\textbf {\bibinfo {volume} {17}},\ \bibinfo {pages} {1114} (\bibinfo {year} {2021})}\BibitemShut {NoStop}%
\bibitem [{\citenamefont {Weston}\ \emph {et~al.}(2020)\citenamefont {Weston}, \citenamefont {Zou}, \citenamefont {Enaldiev}, \citenamefont {Summerfield}, \citenamefont {Clark}, \citenamefont {Z{\'o}lyomi}, \citenamefont {Graham}, \citenamefont {Yelgel}, \citenamefont {Magorrian}, \citenamefont {Zhou} \emph {et~al.}}]{weston2020atomic}%
  \BibitemOpen
  \bibfield  {author} {\bibinfo {author} {\bibfnamefont {A.}~\bibnamefont {Weston}}, \bibinfo {author} {\bibfnamefont {Y.}~\bibnamefont {Zou}}, \bibinfo {author} {\bibfnamefont {V.}~\bibnamefont {Enaldiev}}, \bibinfo {author} {\bibfnamefont {A.}~\bibnamefont {Summerfield}}, \bibinfo {author} {\bibfnamefont {N.}~\bibnamefont {Clark}}, \bibinfo {author} {\bibfnamefont {V.}~\bibnamefont {Z{\'o}lyomi}}, \bibinfo {author} {\bibfnamefont {A.}~\bibnamefont {Graham}}, \bibinfo {author} {\bibfnamefont {C.}~\bibnamefont {Yelgel}}, \bibinfo {author} {\bibfnamefont {S.}~\bibnamefont {Magorrian}}, \bibinfo {author} {\bibfnamefont {M.}~\bibnamefont {Zhou}}, \emph {et~al.},\ }\href@noop {} {\bibfield  {journal} {\bibinfo  {journal} {Nature Nanotechnology}\ }\textbf {\bibinfo {volume} {15}},\ \bibinfo {pages} {592} (\bibinfo {year} {2020})}\BibitemShut {NoStop}%
\bibitem [{\citenamefont {Zeng}\ and\ \citenamefont {MacDonald}(2022)}]{zeng2022strong}%
  \BibitemOpen
  \bibfield  {author} {\bibinfo {author} {\bibfnamefont {Y.}~\bibnamefont {Zeng}}\ and\ \bibinfo {author} {\bibfnamefont {A.~H.}\ \bibnamefont {MacDonald}},\ }\href@noop {} {\bibfield  {journal} {\bibinfo  {journal} {Physical Review B}\ }\textbf {\bibinfo {volume} {106}},\ \bibinfo {pages} {035115} (\bibinfo {year} {2022})}\BibitemShut {NoStop}%
\bibitem [{\citenamefont {Naik}\ \emph {et~al.}(2022)\citenamefont {Naik}, \citenamefont {Regan}, \citenamefont {Zhang}, \citenamefont {Chan}, \citenamefont {Li}, \citenamefont {Wang}, \citenamefont {Yoon}, \citenamefont {Ong}, \citenamefont {Zhao}, \citenamefont {Zhao} \emph {et~al.}}]{naik2022intralayer}%
  \BibitemOpen
  \bibfield  {author} {\bibinfo {author} {\bibfnamefont {M.~H.}\ \bibnamefont {Naik}}, \bibinfo {author} {\bibfnamefont {E.~C.}\ \bibnamefont {Regan}}, \bibinfo {author} {\bibfnamefont {Z.}~\bibnamefont {Zhang}}, \bibinfo {author} {\bibfnamefont {Y.-H.}\ \bibnamefont {Chan}}, \bibinfo {author} {\bibfnamefont {Z.}~\bibnamefont {Li}}, \bibinfo {author} {\bibfnamefont {D.}~\bibnamefont {Wang}}, \bibinfo {author} {\bibfnamefont {Y.}~\bibnamefont {Yoon}}, \bibinfo {author} {\bibfnamefont {C.~S.}\ \bibnamefont {Ong}}, \bibinfo {author} {\bibfnamefont {W.}~\bibnamefont {Zhao}}, \bibinfo {author} {\bibfnamefont {S.}~\bibnamefont {Zhao}}, \emph {et~al.},\ }\href@noop {} {\bibfield  {journal} {\bibinfo  {journal} {Nature}\ }\textbf {\bibinfo {volume} {609}},\ \bibinfo {pages} {52} (\bibinfo {year} {2022})}\BibitemShut {NoStop}%
\bibitem [{\citenamefont {Li}\ \emph {et~al.}(2024)\citenamefont {Li}, \citenamefont {Xiang}, \citenamefont {Naik}, \citenamefont {Kim}, \citenamefont {Li}, \citenamefont {Sailus}, \citenamefont {Banerjee}, \citenamefont {Taniguchi}, \citenamefont {Watanabe}, \citenamefont {Tongay} \emph {et~al.}}]{li2024imaging}%
  \BibitemOpen
  \bibfield  {author} {\bibinfo {author} {\bibfnamefont {H.}~\bibnamefont {Li}}, \bibinfo {author} {\bibfnamefont {Z.}~\bibnamefont {Xiang}}, \bibinfo {author} {\bibfnamefont {M.~H.}\ \bibnamefont {Naik}}, \bibinfo {author} {\bibfnamefont {W.}~\bibnamefont {Kim}}, \bibinfo {author} {\bibfnamefont {Z.}~\bibnamefont {Li}}, \bibinfo {author} {\bibfnamefont {R.}~\bibnamefont {Sailus}}, \bibinfo {author} {\bibfnamefont {R.}~\bibnamefont {Banerjee}}, \bibinfo {author} {\bibfnamefont {T.}~\bibnamefont {Taniguchi}}, \bibinfo {author} {\bibfnamefont {K.}~\bibnamefont {Watanabe}}, \bibinfo {author} {\bibfnamefont {S.}~\bibnamefont {Tongay}}, \emph {et~al.},\ }\href@noop {} {\bibfield  {journal} {\bibinfo  {journal} {Nature materials}\ ,\ \bibinfo {pages} {1}} (\bibinfo {year} {2024})}\BibitemShut {NoStop}%
\bibitem [{SM()}]{SM}%
  \BibitemOpen
  \href@noop {} {\bibinfo {title} {Supplemental material}},\ \bibinfo {note} {see Supplemental Material at [] for additional experimental details, sample characterization, data analysis, and theoretical calculations.}\BibitemShut {Stop}%
\bibitem [{\citenamefont {Carr}\ \emph {et~al.}(2020)\citenamefont {Carr}, \citenamefont {Fang},\ and\ \citenamefont {Kaxiras}}]{carr2020electronic}%
  \BibitemOpen
  \bibfield  {author} {\bibinfo {author} {\bibfnamefont {S.}~\bibnamefont {Carr}}, \bibinfo {author} {\bibfnamefont {S.}~\bibnamefont {Fang}},\ and\ \bibinfo {author} {\bibfnamefont {E.}~\bibnamefont {Kaxiras}},\ }\href@noop {} {\bibfield  {journal} {\bibinfo  {journal} {Nature Reviews Materials}\ }\textbf {\bibinfo {volume} {5}},\ \bibinfo {pages} {748} (\bibinfo {year} {2020})}\BibitemShut {NoStop}%
\bibitem [{\citenamefont {Quan}\ \emph {et~al.}(2021)\citenamefont {Quan}, \citenamefont {Linhart}, \citenamefont {Lin}, \citenamefont {Lee}, \citenamefont {Zhu}, \citenamefont {Wang}, \citenamefont {Hsu}, \citenamefont {Choi}, \citenamefont {Embley}, \citenamefont {Young} \emph {et~al.}}]{quan2021phonon}%
  \BibitemOpen
  \bibfield  {author} {\bibinfo {author} {\bibfnamefont {J.}~\bibnamefont {Quan}}, \bibinfo {author} {\bibfnamefont {L.}~\bibnamefont {Linhart}}, \bibinfo {author} {\bibfnamefont {M.-L.}\ \bibnamefont {Lin}}, \bibinfo {author} {\bibfnamefont {D.}~\bibnamefont {Lee}}, \bibinfo {author} {\bibfnamefont {J.}~\bibnamefont {Zhu}}, \bibinfo {author} {\bibfnamefont {C.-Y.}\ \bibnamefont {Wang}}, \bibinfo {author} {\bibfnamefont {W.-T.}\ \bibnamefont {Hsu}}, \bibinfo {author} {\bibfnamefont {J.}~\bibnamefont {Choi}}, \bibinfo {author} {\bibfnamefont {J.}~\bibnamefont {Embley}}, \bibinfo {author} {\bibfnamefont {C.}~\bibnamefont {Young}}, \emph {et~al.},\ }\href@noop {} {\bibfield  {journal} {\bibinfo  {journal} {Nature Materials}\ }\textbf {\bibinfo {volume} {20}},\ \bibinfo {pages} {1100} (\bibinfo {year} {2021})}\BibitemShut {NoStop}%
\bibitem [{\citenamefont {Puretzky}\ \emph {et~al.}(2016)\citenamefont {Puretzky}, \citenamefont {Liang}, \citenamefont {Li}, \citenamefont {Xiao}, \citenamefont {Sumpter}, \citenamefont {Meunier},\ and\ \citenamefont {Geohegan}}]{puretzky2016twisted}%
  \BibitemOpen
  \bibfield  {author} {\bibinfo {author} {\bibfnamefont {A.~A.}\ \bibnamefont {Puretzky}}, \bibinfo {author} {\bibfnamefont {L.}~\bibnamefont {Liang}}, \bibinfo {author} {\bibfnamefont {X.}~\bibnamefont {Li}}, \bibinfo {author} {\bibfnamefont {K.}~\bibnamefont {Xiao}}, \bibinfo {author} {\bibfnamefont {B.~G.}\ \bibnamefont {Sumpter}}, \bibinfo {author} {\bibfnamefont {V.}~\bibnamefont {Meunier}},\ and\ \bibinfo {author} {\bibfnamefont {D.~B.}\ \bibnamefont {Geohegan}},\ }\href@noop {} {\bibfield  {journal} {\bibinfo  {journal} {ACS Nano}\ }\textbf {\bibinfo {volume} {10}},\ \bibinfo {pages} {2736} (\bibinfo {year} {2016})}\BibitemShut {NoStop}%
\bibitem [{\citenamefont {Helmrich}\ \emph {et~al.}(2021)\citenamefont {Helmrich}, \citenamefont {Sampson}, \citenamefont {Huang}, \citenamefont {Selig}, \citenamefont {Hao}, \citenamefont {Tran}, \citenamefont {Achstein}, \citenamefont {Young}, \citenamefont {Knorr}, \citenamefont {Malic} \emph {et~al.}}]{helmrich2021phonon}%
  \BibitemOpen
  \bibfield  {author} {\bibinfo {author} {\bibfnamefont {S.}~\bibnamefont {Helmrich}}, \bibinfo {author} {\bibfnamefont {K.}~\bibnamefont {Sampson}}, \bibinfo {author} {\bibfnamefont {D.}~\bibnamefont {Huang}}, \bibinfo {author} {\bibfnamefont {M.}~\bibnamefont {Selig}}, \bibinfo {author} {\bibfnamefont {K.}~\bibnamefont {Hao}}, \bibinfo {author} {\bibfnamefont {K.}~\bibnamefont {Tran}}, \bibinfo {author} {\bibfnamefont {A.}~\bibnamefont {Achstein}}, \bibinfo {author} {\bibfnamefont {C.}~\bibnamefont {Young}}, \bibinfo {author} {\bibfnamefont {A.}~\bibnamefont {Knorr}}, \bibinfo {author} {\bibfnamefont {E.}~\bibnamefont {Malic}}, \emph {et~al.},\ }\href@noop {} {\bibfield  {journal} {\bibinfo  {journal} {Physical review letters}\ }\textbf {\bibinfo {volume} {127}},\ \bibinfo {pages} {157403} (\bibinfo {year} {2021})}\BibitemShut {NoStop}%
\bibitem [{\citenamefont {Efimkin}\ \emph {et~al.}(2021)\citenamefont {Efimkin}, \citenamefont {Laird}, \citenamefont {Levinsen}, \citenamefont {Parish},\ and\ \citenamefont {MacDonald}}]{efimkin2021electron}%
  \BibitemOpen
  \bibfield  {author} {\bibinfo {author} {\bibfnamefont {D.~K.}\ \bibnamefont {Efimkin}}, \bibinfo {author} {\bibfnamefont {E.~K.}\ \bibnamefont {Laird}}, \bibinfo {author} {\bibfnamefont {J.}~\bibnamefont {Levinsen}}, \bibinfo {author} {\bibfnamefont {M.~M.}\ \bibnamefont {Parish}},\ and\ \bibinfo {author} {\bibfnamefont {A.~H.}\ \bibnamefont {MacDonald}},\ }\href@noop {} {\bibfield  {journal} {\bibinfo  {journal} {Physical Review B}\ }\textbf {\bibinfo {volume} {103}},\ \bibinfo {pages} {075417} (\bibinfo {year} {2021})}\BibitemShut {NoStop}%
\bibitem [{\citenamefont {Sidler}\ \emph {et~al.}(2017)\citenamefont {Sidler}, \citenamefont {Back}, \citenamefont {Cotlet}, \citenamefont {Srivastava}, \citenamefont {Fink}, \citenamefont {Kroner}, \citenamefont {Demler},\ and\ \citenamefont {Imamoglu}}]{sidler2017fermi}%
  \BibitemOpen
  \bibfield  {author} {\bibinfo {author} {\bibfnamefont {M.}~\bibnamefont {Sidler}}, \bibinfo {author} {\bibfnamefont {P.}~\bibnamefont {Back}}, \bibinfo {author} {\bibfnamefont {O.}~\bibnamefont {Cotlet}}, \bibinfo {author} {\bibfnamefont {A.}~\bibnamefont {Srivastava}}, \bibinfo {author} {\bibfnamefont {T.}~\bibnamefont {Fink}}, \bibinfo {author} {\bibfnamefont {M.}~\bibnamefont {Kroner}}, \bibinfo {author} {\bibfnamefont {E.}~\bibnamefont {Demler}},\ and\ \bibinfo {author} {\bibfnamefont {A.}~\bibnamefont {Imamoglu}},\ }\href@noop {} {\bibfield  {journal} {\bibinfo  {journal} {Nature Physics}\ }\textbf {\bibinfo {volume} {13}},\ \bibinfo {pages} {255} (\bibinfo {year} {2017})}\BibitemShut {NoStop}%
\bibitem [{\citenamefont {Naik}\ \emph {et~al.}(2020)\citenamefont {Naik}, \citenamefont {Kundu}, \citenamefont {Maity},\ and\ \citenamefont {Jain}}]{naik2020origin}%
  \BibitemOpen
  \bibfield  {author} {\bibinfo {author} {\bibfnamefont {M.~H.}\ \bibnamefont {Naik}}, \bibinfo {author} {\bibfnamefont {S.}~\bibnamefont {Kundu}}, \bibinfo {author} {\bibfnamefont {I.}~\bibnamefont {Maity}},\ and\ \bibinfo {author} {\bibfnamefont {M.}~\bibnamefont {Jain}},\ }\href@noop {} {\bibfield  {journal} {\bibinfo  {journal} {Physical Review B}\ }\textbf {\bibinfo {volume} {102}},\ \bibinfo {pages} {075413} (\bibinfo {year} {2020})}\BibitemShut {NoStop}%
\bibitem [{\citenamefont {Gao}\ \emph {et~al.}(2016)\citenamefont {Gao}, \citenamefont {Liang}, \citenamefont {Spataru},\ and\ \citenamefont {Yang}}]{gao2016dynamical}%
  \BibitemOpen
  \bibfield  {author} {\bibinfo {author} {\bibfnamefont {S.}~\bibnamefont {Gao}}, \bibinfo {author} {\bibfnamefont {Y.}~\bibnamefont {Liang}}, \bibinfo {author} {\bibfnamefont {C.~D.}\ \bibnamefont {Spataru}},\ and\ \bibinfo {author} {\bibfnamefont {L.}~\bibnamefont {Yang}},\ }\href@noop {} {\bibfield  {journal} {\bibinfo  {journal} {Nano letters}\ }\textbf {\bibinfo {volume} {16}},\ \bibinfo {pages} {5568} (\bibinfo {year} {2016})}\BibitemShut {NoStop}%
\bibitem [{\citenamefont {F{\"o}rste}\ \emph {et~al.}(2020)\citenamefont {F{\"o}rste}, \citenamefont {Tepliakov}, \citenamefont {Kruchinin}, \citenamefont {Lindlau}, \citenamefont {Funk}, \citenamefont {F{\"o}rg}, \citenamefont {Watanabe}, \citenamefont {Taniguchi}, \citenamefont {Baimuratov},\ and\ \citenamefont {H{\"o}gele}}]{forste2020exciton}%
  \BibitemOpen
  \bibfield  {author} {\bibinfo {author} {\bibfnamefont {J.}~\bibnamefont {F{\"o}rste}}, \bibinfo {author} {\bibfnamefont {N.~V.}\ \bibnamefont {Tepliakov}}, \bibinfo {author} {\bibfnamefont {S.~Y.}\ \bibnamefont {Kruchinin}}, \bibinfo {author} {\bibfnamefont {J.}~\bibnamefont {Lindlau}}, \bibinfo {author} {\bibfnamefont {V.}~\bibnamefont {Funk}}, \bibinfo {author} {\bibfnamefont {M.}~\bibnamefont {F{\"o}rg}}, \bibinfo {author} {\bibfnamefont {K.}~\bibnamefont {Watanabe}}, \bibinfo {author} {\bibfnamefont {T.}~\bibnamefont {Taniguchi}}, \bibinfo {author} {\bibfnamefont {A.~S.}\ \bibnamefont {Baimuratov}},\ and\ \bibinfo {author} {\bibfnamefont {A.}~\bibnamefont {H{\"o}gele}},\ }\href@noop {} {\bibfield  {journal} {\bibinfo  {journal} {Nature Communications}\ }\textbf {\bibinfo {volume} {11}},\ \bibinfo {pages} {4539} (\bibinfo {year} {2020})}\BibitemShut {NoStop}%
\bibitem [{\citenamefont {Smole{\'n}ski}\ \emph {et~al.}(2019)\citenamefont {Smole{\'n}ski}, \citenamefont {Cotlet}, \citenamefont {Popert}, \citenamefont {Back}, \citenamefont {Shimazaki}, \citenamefont {Kn{\"u}ppel}, \citenamefont {Dietler}, \citenamefont {Taniguchi}, \citenamefont {Watanabe}, \citenamefont {Kroner} \emph {et~al.}}]{smolenski2019interaction}%
  \BibitemOpen
  \bibfield  {author} {\bibinfo {author} {\bibfnamefont {T.}~\bibnamefont {Smole{\'n}ski}}, \bibinfo {author} {\bibfnamefont {O.}~\bibnamefont {Cotlet}}, \bibinfo {author} {\bibfnamefont {A.}~\bibnamefont {Popert}}, \bibinfo {author} {\bibfnamefont {P.}~\bibnamefont {Back}}, \bibinfo {author} {\bibfnamefont {Y.}~\bibnamefont {Shimazaki}}, \bibinfo {author} {\bibfnamefont {P.}~\bibnamefont {Kn{\"u}ppel}}, \bibinfo {author} {\bibfnamefont {N.}~\bibnamefont {Dietler}}, \bibinfo {author} {\bibfnamefont {T.}~\bibnamefont {Taniguchi}}, \bibinfo {author} {\bibfnamefont {K.}~\bibnamefont {Watanabe}}, \bibinfo {author} {\bibfnamefont {M.}~\bibnamefont {Kroner}}, \emph {et~al.},\ }\href@noop {} {\bibfield  {journal} {\bibinfo  {journal} {Physical Review Letters}\ }\textbf {\bibinfo {volume} {123}},\ \bibinfo {pages} {097403} (\bibinfo {year} {2019})}\BibitemShut {NoStop}%
\bibitem [{\citenamefont {Campbell}\ \emph {et~al.}(2024)\citenamefont {Campbell}, \citenamefont {Vitale}, \citenamefont {Brotons-Gisbert}, \citenamefont {Baek}, \citenamefont {Borel}, \citenamefont {Ivanova}, \citenamefont {Taniguchi}, \citenamefont {Watanabe}, \citenamefont {Lischner},\ and\ \citenamefont {Gerardot}}]{campbell2024interplay}%
  \BibitemOpen
  \bibfield  {author} {\bibinfo {author} {\bibfnamefont {A.~J.}\ \bibnamefont {Campbell}}, \bibinfo {author} {\bibfnamefont {V.}~\bibnamefont {Vitale}}, \bibinfo {author} {\bibfnamefont {M.}~\bibnamefont {Brotons-Gisbert}}, \bibinfo {author} {\bibfnamefont {H.}~\bibnamefont {Baek}}, \bibinfo {author} {\bibfnamefont {A.}~\bibnamefont {Borel}}, \bibinfo {author} {\bibfnamefont {T.~V.}\ \bibnamefont {Ivanova}}, \bibinfo {author} {\bibfnamefont {T.}~\bibnamefont {Taniguchi}}, \bibinfo {author} {\bibfnamefont {K.}~\bibnamefont {Watanabe}}, \bibinfo {author} {\bibfnamefont {J.}~\bibnamefont {Lischner}},\ and\ \bibinfo {author} {\bibfnamefont {B.~D.}\ \bibnamefont {Gerardot}},\ }\href@noop {} {\bibfield  {journal} {\bibinfo  {journal} {Nature Physics}\ ,\ \bibinfo {pages} {1}} (\bibinfo {year} {2024})}\BibitemShut {NoStop}%
\bibitem [{\citenamefont {Hybertsen}\ and\ \citenamefont {Louie}(1986{\natexlab{a}})}]{hybertsen1986electron}%
  \BibitemOpen
  \bibfield  {author} {\bibinfo {author} {\bibfnamefont {M.~S.}\ \bibnamefont {Hybertsen}}\ and\ \bibinfo {author} {\bibfnamefont {S.~G.}\ \bibnamefont {Louie}},\ }\href@noop {} {\bibfield  {journal} {\bibinfo  {journal} {Physical Review B}\ }\textbf {\bibinfo {volume} {34}},\ \bibinfo {pages} {5390} (\bibinfo {year} {1986}{\natexlab{a}})}\BibitemShut {NoStop}%
\bibitem [{\citenamefont {Xian}\ \emph {et~al.}(2021)\citenamefont {Xian}, \citenamefont {Claassen}, \citenamefont {Kiese}, \citenamefont {Scherer}, \citenamefont {Trebst}, \citenamefont {Kennes},\ and\ \citenamefont {Rubio}}]{xian2021realization}%
  \BibitemOpen
  \bibfield  {author} {\bibinfo {author} {\bibfnamefont {L.}~\bibnamefont {Xian}}, \bibinfo {author} {\bibfnamefont {M.}~\bibnamefont {Claassen}}, \bibinfo {author} {\bibfnamefont {D.}~\bibnamefont {Kiese}}, \bibinfo {author} {\bibfnamefont {M.~M.}\ \bibnamefont {Scherer}}, \bibinfo {author} {\bibfnamefont {S.}~\bibnamefont {Trebst}}, \bibinfo {author} {\bibfnamefont {D.~M.}\ \bibnamefont {Kennes}},\ and\ \bibinfo {author} {\bibfnamefont {A.}~\bibnamefont {Rubio}},\ }\href@noop {} {\bibfield  {journal} {\bibinfo  {journal} {Nature Communications}\ }\textbf {\bibinfo {volume} {12}},\ \bibinfo {pages} {5644} (\bibinfo {year} {2021})}\BibitemShut {NoStop}%
\bibitem [{\citenamefont {Qiu}\ \emph {et~al.}(2013)\citenamefont {Qiu}, \citenamefont {Da~Jornada},\ and\ \citenamefont {Louie}}]{qiu2013optical}%
  \BibitemOpen
  \bibfield  {author} {\bibinfo {author} {\bibfnamefont {D.~Y.}\ \bibnamefont {Qiu}}, \bibinfo {author} {\bibfnamefont {F.~H.}\ \bibnamefont {Da~Jornada}},\ and\ \bibinfo {author} {\bibfnamefont {S.~G.}\ \bibnamefont {Louie}},\ }\href@noop {} {\bibfield  {journal} {\bibinfo  {journal} {Physical review letters}\ }\textbf {\bibinfo {volume} {111}},\ \bibinfo {pages} {216805} (\bibinfo {year} {2013})}\BibitemShut {NoStop}%
\bibitem [{\citenamefont {Rohlfing}\ and\ \citenamefont {Louie}(2000{\natexlab{a}})}]{rohlfing2000electron}%
  \BibitemOpen
  \bibfield  {author} {\bibinfo {author} {\bibfnamefont {M.}~\bibnamefont {Rohlfing}}\ and\ \bibinfo {author} {\bibfnamefont {S.~G.}\ \bibnamefont {Louie}},\ }\href@noop {} {\bibfield  {journal} {\bibinfo  {journal} {Physical Review B}\ }\textbf {\bibinfo {volume} {62}},\ \bibinfo {pages} {4927} (\bibinfo {year} {2000}{\natexlab{a}})}\BibitemShut {NoStop}%
\bibitem [{\citenamefont {Susarla}\ \emph {et~al.}(2022)\citenamefont {Susarla}, \citenamefont {Naik}, \citenamefont {Blach}, \citenamefont {Zipfel}, \citenamefont {Taniguchi}, \citenamefont {Watanabe}, \citenamefont {Huang}, \citenamefont {Ramesh}, \citenamefont {da~Jornada}, \citenamefont {Louie} \emph {et~al.}}]{susarla2022hyperspectral}%
  \BibitemOpen
  \bibfield  {author} {\bibinfo {author} {\bibfnamefont {S.}~\bibnamefont {Susarla}}, \bibinfo {author} {\bibfnamefont {M.~H.}\ \bibnamefont {Naik}}, \bibinfo {author} {\bibfnamefont {D.~D.}\ \bibnamefont {Blach}}, \bibinfo {author} {\bibfnamefont {J.}~\bibnamefont {Zipfel}}, \bibinfo {author} {\bibfnamefont {T.}~\bibnamefont {Taniguchi}}, \bibinfo {author} {\bibfnamefont {K.}~\bibnamefont {Watanabe}}, \bibinfo {author} {\bibfnamefont {L.}~\bibnamefont {Huang}}, \bibinfo {author} {\bibfnamefont {R.}~\bibnamefont {Ramesh}}, \bibinfo {author} {\bibfnamefont {F.~H.}\ \bibnamefont {da~Jornada}}, \bibinfo {author} {\bibfnamefont {S.~G.}\ \bibnamefont {Louie}}, \emph {et~al.},\ }\href@noop {} {\bibfield  {journal} {\bibinfo  {journal} {Science}\ }\textbf {\bibinfo {volume} {378}},\ \bibinfo {pages} {1235} (\bibinfo {year} {2022})}\BibitemShut {NoStop}%
\bibitem [{\citenamefont {Miao}\ \emph {et~al.}(2021)\citenamefont {Miao}, \citenamefont {Wang}, \citenamefont {Huang}, \citenamefont {Chen}, \citenamefont {Lian}, \citenamefont {Wang}, \citenamefont {Blei}, \citenamefont {Taniguchi}, \citenamefont {Watanabe}, \citenamefont {Tongay} \emph {et~al.}}]{miao2021strong}%
  \BibitemOpen
  \bibfield  {author} {\bibinfo {author} {\bibfnamefont {S.}~\bibnamefont {Miao}}, \bibinfo {author} {\bibfnamefont {T.}~\bibnamefont {Wang}}, \bibinfo {author} {\bibfnamefont {X.}~\bibnamefont {Huang}}, \bibinfo {author} {\bibfnamefont {D.}~\bibnamefont {Chen}}, \bibinfo {author} {\bibfnamefont {Z.}~\bibnamefont {Lian}}, \bibinfo {author} {\bibfnamefont {C.}~\bibnamefont {Wang}}, \bibinfo {author} {\bibfnamefont {M.}~\bibnamefont {Blei}}, \bibinfo {author} {\bibfnamefont {T.}~\bibnamefont {Taniguchi}}, \bibinfo {author} {\bibfnamefont {K.}~\bibnamefont {Watanabe}}, \bibinfo {author} {\bibfnamefont {S.}~\bibnamefont {Tongay}}, \emph {et~al.},\ }\href@noop {} {\bibfield  {journal} {\bibinfo  {journal} {Nature communications}\ }\textbf {\bibinfo {volume} {12}},\ \bibinfo {pages} {3608} (\bibinfo {year} {2021})}\BibitemShut {NoStop}%
\bibitem [{\citenamefont {Wang}\ \emph {et~al.}(2023)\citenamefont {Wang}, \citenamefont {Zhang}, \citenamefont {Zhu}, \citenamefont {Park}, \citenamefont {Wang}, \citenamefont {Wang}, \citenamefont {Holtzmann}, \citenamefont {Taniguchi}, \citenamefont {Watanabe}, \citenamefont {Yan} \emph {et~al.}}]{wang2023intercell}%
  \BibitemOpen
  \bibfield  {author} {\bibinfo {author} {\bibfnamefont {X.}~\bibnamefont {Wang}}, \bibinfo {author} {\bibfnamefont {X.}~\bibnamefont {Zhang}}, \bibinfo {author} {\bibfnamefont {J.}~\bibnamefont {Zhu}}, \bibinfo {author} {\bibfnamefont {H.}~\bibnamefont {Park}}, \bibinfo {author} {\bibfnamefont {Y.}~\bibnamefont {Wang}}, \bibinfo {author} {\bibfnamefont {C.}~\bibnamefont {Wang}}, \bibinfo {author} {\bibfnamefont {W.~G.}\ \bibnamefont {Holtzmann}}, \bibinfo {author} {\bibfnamefont {T.}~\bibnamefont {Taniguchi}}, \bibinfo {author} {\bibfnamefont {K.}~\bibnamefont {Watanabe}}, \bibinfo {author} {\bibfnamefont {J.}~\bibnamefont {Yan}}, \emph {et~al.},\ }\href@noop {} {\bibfield  {journal} {\bibinfo  {journal} {Nature Materials}\ ,\ \bibinfo {pages} {1}} (\bibinfo {year} {2023})}\BibitemShut {NoStop}%
\bibitem [{\citenamefont {Chu}\ \emph {et~al.}(2020)\citenamefont {Chu}, \citenamefont {Regan}, \citenamefont {Ma}, \citenamefont {Wang}, \citenamefont {Xu}, \citenamefont {Utama}, \citenamefont {Yumigeta}, \citenamefont {Blei}, \citenamefont {Watanabe}, \citenamefont {Taniguchi} \emph {et~al.}}]{chu2020nanoscale}%
  \BibitemOpen
  \bibfield  {author} {\bibinfo {author} {\bibfnamefont {Z.}~\bibnamefont {Chu}}, \bibinfo {author} {\bibfnamefont {E.~C.}\ \bibnamefont {Regan}}, \bibinfo {author} {\bibfnamefont {X.}~\bibnamefont {Ma}}, \bibinfo {author} {\bibfnamefont {D.}~\bibnamefont {Wang}}, \bibinfo {author} {\bibfnamefont {Z.}~\bibnamefont {Xu}}, \bibinfo {author} {\bibfnamefont {M.~I.~B.}\ \bibnamefont {Utama}}, \bibinfo {author} {\bibfnamefont {K.}~\bibnamefont {Yumigeta}}, \bibinfo {author} {\bibfnamefont {M.}~\bibnamefont {Blei}}, \bibinfo {author} {\bibfnamefont {K.}~\bibnamefont {Watanabe}}, \bibinfo {author} {\bibfnamefont {T.}~\bibnamefont {Taniguchi}}, \emph {et~al.},\ }\href@noop {} {\bibfield  {journal} {\bibinfo  {journal} {Physical review letters}\ }\textbf {\bibinfo {volume} {125}},\ \bibinfo {pages} {186803} (\bibinfo {year} {2020})}\BibitemShut {NoStop}%
\bibitem [{\citenamefont {Huang}\ \emph {et~al.}(2021)\citenamefont {Huang}, \citenamefont {Wang}, \citenamefont {Miao}, \citenamefont {Wang}, \citenamefont {Li}, \citenamefont {Lian}, \citenamefont {Taniguchi}, \citenamefont {Watanabe}, \citenamefont {Okamoto}, \citenamefont {Xiao} \emph {et~al.}}]{huang2021correlated}%
  \BibitemOpen
  \bibfield  {author} {\bibinfo {author} {\bibfnamefont {X.}~\bibnamefont {Huang}}, \bibinfo {author} {\bibfnamefont {T.}~\bibnamefont {Wang}}, \bibinfo {author} {\bibfnamefont {S.}~\bibnamefont {Miao}}, \bibinfo {author} {\bibfnamefont {C.}~\bibnamefont {Wang}}, \bibinfo {author} {\bibfnamefont {Z.}~\bibnamefont {Li}}, \bibinfo {author} {\bibfnamefont {Z.}~\bibnamefont {Lian}}, \bibinfo {author} {\bibfnamefont {T.}~\bibnamefont {Taniguchi}}, \bibinfo {author} {\bibfnamefont {K.}~\bibnamefont {Watanabe}}, \bibinfo {author} {\bibfnamefont {S.}~\bibnamefont {Okamoto}}, \bibinfo {author} {\bibfnamefont {D.}~\bibnamefont {Xiao}}, \emph {et~al.},\ }\href@noop {} {\bibfield  {journal} {\bibinfo  {journal} {Nature Physics}\ }\textbf {\bibinfo {volume} {17}},\ \bibinfo {pages} {715} (\bibinfo {year} {2021})}\BibitemShut {NoStop}%
\bibitem [{\citenamefont {Wang}\ \emph {et~al.}(2021)\citenamefont {Wang}, \citenamefont {Zhu}, \citenamefont {Seyler}, \citenamefont {Rivera}, \citenamefont {Zheng}, \citenamefont {Wang}, \citenamefont {He}, \citenamefont {Taniguchi}, \citenamefont {Watanabe}, \citenamefont {Yan} \emph {et~al.}}]{wang2021moire}%
  \BibitemOpen
  \bibfield  {author} {\bibinfo {author} {\bibfnamefont {X.}~\bibnamefont {Wang}}, \bibinfo {author} {\bibfnamefont {J.}~\bibnamefont {Zhu}}, \bibinfo {author} {\bibfnamefont {K.~L.}\ \bibnamefont {Seyler}}, \bibinfo {author} {\bibfnamefont {P.}~\bibnamefont {Rivera}}, \bibinfo {author} {\bibfnamefont {H.}~\bibnamefont {Zheng}}, \bibinfo {author} {\bibfnamefont {Y.}~\bibnamefont {Wang}}, \bibinfo {author} {\bibfnamefont {M.}~\bibnamefont {He}}, \bibinfo {author} {\bibfnamefont {T.}~\bibnamefont {Taniguchi}}, \bibinfo {author} {\bibfnamefont {K.}~\bibnamefont {Watanabe}}, \bibinfo {author} {\bibfnamefont {J.}~\bibnamefont {Yan}}, \emph {et~al.},\ }\href@noop {} {\bibfield  {journal} {\bibinfo  {journal} {Nature Nanotechnology}\ }\textbf {\bibinfo {volume} {16}},\ \bibinfo {pages} {1208} (\bibinfo {year} {2021})}\BibitemShut {NoStop}%
\bibitem [{\citenamefont {Liu}\ \emph {et~al.}(2021)\citenamefont {Liu}, \citenamefont {Barr{\'e}}, \citenamefont {van Baren}, \citenamefont {Wilson}, \citenamefont {Taniguchi}, \citenamefont {Watanabe}, \citenamefont {Cui}, \citenamefont {Gabor}, \citenamefont {Heinz}, \citenamefont {Chang} \emph {et~al.}}]{liu2021signatures}%
  \BibitemOpen
  \bibfield  {author} {\bibinfo {author} {\bibfnamefont {E.}~\bibnamefont {Liu}}, \bibinfo {author} {\bibfnamefont {E.}~\bibnamefont {Barr{\'e}}}, \bibinfo {author} {\bibfnamefont {J.}~\bibnamefont {van Baren}}, \bibinfo {author} {\bibfnamefont {M.}~\bibnamefont {Wilson}}, \bibinfo {author} {\bibfnamefont {T.}~\bibnamefont {Taniguchi}}, \bibinfo {author} {\bibfnamefont {K.}~\bibnamefont {Watanabe}}, \bibinfo {author} {\bibfnamefont {Y.-T.}\ \bibnamefont {Cui}}, \bibinfo {author} {\bibfnamefont {N.~M.}\ \bibnamefont {Gabor}}, \bibinfo {author} {\bibfnamefont {T.~F.}\ \bibnamefont {Heinz}}, \bibinfo {author} {\bibfnamefont {Y.-C.}\ \bibnamefont {Chang}}, \emph {et~al.},\ }\href@noop {} {\bibfield  {journal} {\bibinfo  {journal} {Nature}\ }\textbf {\bibinfo {volume} {594}},\ \bibinfo {pages} {46} (\bibinfo {year} {2021})}\BibitemShut {NoStop}%
\bibitem [{\citenamefont {Brotons-Gisbert}\ \emph {et~al.}(2021)\citenamefont {Brotons-Gisbert}, \citenamefont {Baek}, \citenamefont {Campbell}, \citenamefont {Watanabe}, \citenamefont {Taniguchi},\ and\ \citenamefont {Gerardot}}]{brotons2021moire}%
  \BibitemOpen
  \bibfield  {author} {\bibinfo {author} {\bibfnamefont {M.}~\bibnamefont {Brotons-Gisbert}}, \bibinfo {author} {\bibfnamefont {H.}~\bibnamefont {Baek}}, \bibinfo {author} {\bibfnamefont {A.}~\bibnamefont {Campbell}}, \bibinfo {author} {\bibfnamefont {K.}~\bibnamefont {Watanabe}}, \bibinfo {author} {\bibfnamefont {T.}~\bibnamefont {Taniguchi}},\ and\ \bibinfo {author} {\bibfnamefont {B.~D.}\ \bibnamefont {Gerardot}},\ }\href@noop {} {\bibfield  {journal} {\bibinfo  {journal} {Physical Review X}\ }\textbf {\bibinfo {volume} {11}},\ \bibinfo {pages} {031033} (\bibinfo {year} {2021})}\BibitemShut {NoStop}%
\bibitem [{\citenamefont {Kroutvar}\ \emph {et~al.}(2004)\citenamefont {Kroutvar}, \citenamefont {Ducommun}, \citenamefont {Heiss}, \citenamefont {Bichler}, \citenamefont {Schuh}, \citenamefont {Abstreiter},\ and\ \citenamefont {Finley}}]{kroutvar2004optically}%
  \BibitemOpen
  \bibfield  {author} {\bibinfo {author} {\bibfnamefont {M.}~\bibnamefont {Kroutvar}}, \bibinfo {author} {\bibfnamefont {Y.}~\bibnamefont {Ducommun}}, \bibinfo {author} {\bibfnamefont {D.}~\bibnamefont {Heiss}}, \bibinfo {author} {\bibfnamefont {M.}~\bibnamefont {Bichler}}, \bibinfo {author} {\bibfnamefont {D.}~\bibnamefont {Schuh}}, \bibinfo {author} {\bibfnamefont {G.}~\bibnamefont {Abstreiter}},\ and\ \bibinfo {author} {\bibfnamefont {J.~J.}\ \bibnamefont {Finley}},\ }\href@noop {} {\bibfield  {journal} {\bibinfo  {journal} {Nature}\ }\textbf {\bibinfo {volume} {432}},\ \bibinfo {pages} {81} (\bibinfo {year} {2004})}\BibitemShut {NoStop}%
\bibitem [{\citenamefont {Liu}(2026)}]{liu2026figshare}%
  \BibitemOpen
  \bibfield  {author} {\bibinfo {author} {\bibfnamefont {Z.}~\bibnamefont {Liu}},\ }\href {https://doi.org/10.6084/m9.figshare.32655012.v1} {\bibinfo {title} {Raw data for {PRL} paper ``emergent trion resonance driven by lattice reconstruction in a moir{'e} superlattice''}} (\bibinfo {year} {2026}),\ \bibinfo {note} {dataset}\BibitemShut {NoStop}%
\bibitem [{\citenamefont {Lim}\ \emph {et~al.}(2023)\citenamefont {Lim}, \citenamefont {Kim}, \citenamefont {Choi}, \citenamefont {Taniguchi}, \citenamefont {Watanabe}, \citenamefont {Choi},\ and\ \citenamefont {Cheong}}]{lim2023modulation}%
  \BibitemOpen
  \bibfield  {author} {\bibinfo {author} {\bibfnamefont {S.~Y.}\ \bibnamefont {Lim}}, \bibinfo {author} {\bibfnamefont {H.-g.}\ \bibnamefont {Kim}}, \bibinfo {author} {\bibfnamefont {Y.~W.}\ \bibnamefont {Choi}}, \bibinfo {author} {\bibfnamefont {T.}~\bibnamefont {Taniguchi}}, \bibinfo {author} {\bibfnamefont {K.}~\bibnamefont {Watanabe}}, \bibinfo {author} {\bibfnamefont {H.~J.}\ \bibnamefont {Choi}},\ and\ \bibinfo {author} {\bibfnamefont {H.}~\bibnamefont {Cheong}},\ }\href@noop {} {\bibfield  {journal} {\bibinfo  {journal} {Acs Nano}\ }\textbf {\bibinfo {volume} {17}},\ \bibinfo {pages} {13938} (\bibinfo {year} {2023})}\BibitemShut {NoStop}%
\bibitem [{\citenamefont {Giannozzi}\ \emph {et~al.}(2009)\citenamefont {Giannozzi}, \citenamefont {Baroni}, \citenamefont {Bonini}, \citenamefont {Calandra}, \citenamefont {Car}, \citenamefont {Cavazzoni}, \citenamefont {Ceresoli}, \citenamefont {Chiarotti}, \citenamefont {Cococcioni}, \citenamefont {Dabo} \emph {et~al.}}]{giannozzi2009quantum}%
  \BibitemOpen
  \bibfield  {author} {\bibinfo {author} {\bibfnamefont {P.}~\bibnamefont {Giannozzi}}, \bibinfo {author} {\bibfnamefont {S.}~\bibnamefont {Baroni}}, \bibinfo {author} {\bibfnamefont {N.}~\bibnamefont {Bonini}}, \bibinfo {author} {\bibfnamefont {M.}~\bibnamefont {Calandra}}, \bibinfo {author} {\bibfnamefont {R.}~\bibnamefont {Car}}, \bibinfo {author} {\bibfnamefont {C.}~\bibnamefont {Cavazzoni}}, \bibinfo {author} {\bibfnamefont {D.}~\bibnamefont {Ceresoli}}, \bibinfo {author} {\bibfnamefont {G.~L.}\ \bibnamefont {Chiarotti}}, \bibinfo {author} {\bibfnamefont {M.}~\bibnamefont {Cococcioni}}, \bibinfo {author} {\bibfnamefont {I.}~\bibnamefont {Dabo}}, \emph {et~al.},\ }\href@noop {} {\bibfield  {journal} {\bibinfo  {journal} {Journal of Physics: Condensed matter}\ }\textbf {\bibinfo {volume} {21}},\ \bibinfo {pages} {395502} (\bibinfo {year} {2009})}\BibitemShut {NoStop}%
\bibitem [{\citenamefont {Hybertsen}\ and\ \citenamefont {Louie}(1986{\natexlab{b}})}]{PhysRevB.34.5390}%
  \BibitemOpen
  \bibfield  {author} {\bibinfo {author} {\bibfnamefont {M.~S.}\ \bibnamefont {Hybertsen}}\ and\ \bibinfo {author} {\bibfnamefont {S.~G.}\ \bibnamefont {Louie}},\ }\href {https://doi.org/10.1103/PhysRevB.34.5390} {\bibfield  {journal} {\bibinfo  {journal} {Physical Review B}\ }\textbf {\bibinfo {volume} {34}},\ \bibinfo {pages} {5390} (\bibinfo {year} {1986}{\natexlab{b}})}\BibitemShut {NoStop}%
\bibitem [{\citenamefont {Rohlfing}\ and\ \citenamefont {Louie}(2000{\natexlab{b}})}]{PhysRevB.62.4927}%
  \BibitemOpen
  \bibfield  {author} {\bibinfo {author} {\bibfnamefont {M.}~\bibnamefont {Rohlfing}}\ and\ \bibinfo {author} {\bibfnamefont {S.~G.}\ \bibnamefont {Louie}},\ }\href {https://doi.org/10.1103/PhysRevB.62.4927} {\bibfield  {journal} {\bibinfo  {journal} {Physical Review B}\ }\textbf {\bibinfo {volume} {62}},\ \bibinfo {pages} {4927} (\bibinfo {year} {2000}{\natexlab{b}})}\BibitemShut {NoStop}%
\bibitem [{\citenamefont {Van~Tuan}\ \emph {et~al.}(2019)\citenamefont {Van~Tuan}, \citenamefont {Scharf}, \citenamefont {Wang}, \citenamefont {Shan}, \citenamefont {Mak}, \citenamefont {{\v{Z}}uti{\'c}},\ and\ \citenamefont {Dery}}]{van2019probing}%
  \BibitemOpen
  \bibfield  {author} {\bibinfo {author} {\bibfnamefont {D.}~\bibnamefont {Van~Tuan}}, \bibinfo {author} {\bibfnamefont {B.}~\bibnamefont {Scharf}}, \bibinfo {author} {\bibfnamefont {Z.}~\bibnamefont {Wang}}, \bibinfo {author} {\bibfnamefont {J.}~\bibnamefont {Shan}}, \bibinfo {author} {\bibfnamefont {K.~F.}\ \bibnamefont {Mak}}, \bibinfo {author} {\bibfnamefont {I.}~\bibnamefont {{\v{Z}}uti{\'c}}},\ and\ \bibinfo {author} {\bibfnamefont {H.}~\bibnamefont {Dery}},\ }\href@noop {} {\bibfield  {journal} {\bibinfo  {journal} {Physical Review B}\ }\textbf {\bibinfo {volume} {99}},\ \bibinfo {pages} {085301} (\bibinfo {year} {2019})}\BibitemShut {NoStop}%
\bibitem [{\citenamefont {Thompson}\ \emph {et~al.}(2022)\citenamefont {Thompson}, \citenamefont {Aktulga}, \citenamefont {Berger}, \citenamefont {Bolintineanu}, \citenamefont {Brown}, \citenamefont {Crozier}, \citenamefont {In't~Veld}, \citenamefont {Kohlmeyer}, \citenamefont {Moore}, \citenamefont {Nguyen} \emph {et~al.}}]{thompson2022lammps}%
  \BibitemOpen
  \bibfield  {author} {\bibinfo {author} {\bibfnamefont {A.~P.}\ \bibnamefont {Thompson}}, \bibinfo {author} {\bibfnamefont {H.~M.}\ \bibnamefont {Aktulga}}, \bibinfo {author} {\bibfnamefont {R.}~\bibnamefont {Berger}}, \bibinfo {author} {\bibfnamefont {D.~S.}\ \bibnamefont {Bolintineanu}}, \bibinfo {author} {\bibfnamefont {W.~M.}\ \bibnamefont {Brown}}, \bibinfo {author} {\bibfnamefont {P.~S.}\ \bibnamefont {Crozier}}, \bibinfo {author} {\bibfnamefont {P.~J.}\ \bibnamefont {In't~Veld}}, \bibinfo {author} {\bibfnamefont {A.}~\bibnamefont {Kohlmeyer}}, \bibinfo {author} {\bibfnamefont {S.~G.}\ \bibnamefont {Moore}}, \bibinfo {author} {\bibfnamefont {T.~D.}\ \bibnamefont {Nguyen}}, \emph {et~al.},\ }\href@noop {} {\bibfield  {journal} {\bibinfo  {journal} {Computer Physics Communications}\ }\textbf {\bibinfo {volume} {271}},\ \bibinfo {pages} {108171} (\bibinfo {year} {2022})}\BibitemShut {NoStop}%
\bibitem [{\citenamefont {Jiang}(2017)}]{jiang2017handbook}%
  \BibitemOpen
  \bibfield  {author} {\bibinfo {author} {\bibfnamefont {J.-W.}\ \bibnamefont {Jiang}},\ }\href@noop {} {\emph {\bibinfo {title} {Handbook of Stillinger-Weber Potential Parameters for Two-Dimensional Atomic Crystals}}}\ (\bibinfo  {publisher} {BoD--Books on Demand},\ \bibinfo {year} {2017})\BibitemShut {NoStop}%
\bibitem [{\citenamefont {Naik}\ \emph {et~al.}(2019)\citenamefont {Naik}, \citenamefont {Maity}, \citenamefont {Maiti},\ and\ \citenamefont {Jain}}]{naik2019kolmogorov}%
  \BibitemOpen
  \bibfield  {author} {\bibinfo {author} {\bibfnamefont {M.~H.}\ \bibnamefont {Naik}}, \bibinfo {author} {\bibfnamefont {I.}~\bibnamefont {Maity}}, \bibinfo {author} {\bibfnamefont {P.~K.}\ \bibnamefont {Maiti}},\ and\ \bibinfo {author} {\bibfnamefont {M.}~\bibnamefont {Jain}},\ }\href@noop {} {\bibfield  {journal} {\bibinfo  {journal} {The Journal of Physical Chemistry C}\ }\textbf {\bibinfo {volume} {123}},\ \bibinfo {pages} {9770} (\bibinfo {year} {2019})}\BibitemShut {NoStop}%
\bibitem [{\citenamefont {Song}\ and\ \citenamefont {Yang}(2017)}]{PhysRevB.96.235441}%
  \BibitemOpen
  \bibfield  {author} {\bibinfo {author} {\bibfnamefont {W.}~\bibnamefont {Song}}\ and\ \bibinfo {author} {\bibfnamefont {L.}~\bibnamefont {Yang}},\ }\href {https://doi.org/10.1103/PhysRevB.96.235441} {\bibfield  {journal} {\bibinfo  {journal} {Phys. Rev. B}\ }\textbf {\bibinfo {volume} {96}},\ \bibinfo {pages} {235441} (\bibinfo {year} {2017})}\BibitemShut {NoStop}%
\bibitem [{\citenamefont {Geick}\ \emph {et~al.}(1966)\citenamefont {Geick}, \citenamefont {Perry},\ and\ \citenamefont {Rupprecht}}]{geick1966normal}%
  \BibitemOpen
  \bibfield  {author} {\bibinfo {author} {\bibfnamefont {R.}~\bibnamefont {Geick}}, \bibinfo {author} {\bibfnamefont {C.}~\bibnamefont {Perry}},\ and\ \bibinfo {author} {\bibfnamefont {G.}~\bibnamefont {Rupprecht}},\ }\href@noop {} {\bibfield  {journal} {\bibinfo  {journal} {Physical Review}\ }\textbf {\bibinfo {volume} {146}},\ \bibinfo {pages} {543} (\bibinfo {year} {1966})}\BibitemShut {NoStop}%
\bibitem [{\citenamefont {Zhang}\ \emph {et~al.}(2014)\citenamefont {Zhang}, \citenamefont {Wang}, \citenamefont {Chan}, \citenamefont {Manolatou},\ and\ \citenamefont {Rana}}]{zhang2014absorption}%
  \BibitemOpen
  \bibfield  {author} {\bibinfo {author} {\bibfnamefont {C.}~\bibnamefont {Zhang}}, \bibinfo {author} {\bibfnamefont {H.}~\bibnamefont {Wang}}, \bibinfo {author} {\bibfnamefont {W.}~\bibnamefont {Chan}}, \bibinfo {author} {\bibfnamefont {C.}~\bibnamefont {Manolatou}},\ and\ \bibinfo {author} {\bibfnamefont {F.}~\bibnamefont {Rana}},\ }\href@noop {} {\bibfield  {journal} {\bibinfo  {journal} {Physical Review B}\ }\textbf {\bibinfo {volume} {89}},\ \bibinfo {pages} {205436} (\bibinfo {year} {2014})}\BibitemShut {NoStop}%
\bibitem [{\citenamefont {Born}\ and\ \citenamefont {Wolf}(2013)}]{born2013principles}%
  \BibitemOpen
  \bibfield  {author} {\bibinfo {author} {\bibfnamefont {M.}~\bibnamefont {Born}}\ and\ \bibinfo {author} {\bibfnamefont {E.}~\bibnamefont {Wolf}},\ }\href@noop {} {\emph {\bibinfo {title} {Principles of optics: electromagnetic theory of propagation, interference and diffraction of light}}}\ (\bibinfo  {publisher} {Elsevier},\ \bibinfo {year} {2013})\BibitemShut {NoStop}%
\end{thebibliography}
\end{document}